\documentstyle[12pt,aaspp4]{article}

\lefthead{Morris et al.}
\righthead{Galaxy Evolution in MS1621.5+2640}

\begin{document}

\title{Galaxy Evolution in the z=0.4274 Cluster
MS1621.5+2640\footnote{Based on Observations taken at the
Canada-France-Hawaii Telescope, which is operated by the NRC of
Canada, CNRS of France and the University of Hawaii.}}

\author{Simon L. Morris and J. B. Hutchings}

\affil{Dominion Astrophysical Observatory, National Research Council,\\
 5071 West Saanich Road, Victoria, B.C., V8X 4M6, Canada,\\ 
 e-mail: Simon.Morris@hia.nrc.ca, John.Hutchings@hia.nrc.ca}

\author{R. G. Carlberg and H. K. C. Yee}

\affil{Department of Astronomy, University of Toronto, Toronto, \\
 Ontario M5S 1A7, Canada, \\
 email: carlberg@moonray.astro.utoronto.ca, hyee@makalu.astro.utoronto.ca}

\author{Erica Ellingson}

\affil{CASA, University of Colorado, Boulder, CO 80309-0389, \\
 email: e.elling@pisco.colorado.edu}

\author{Mike L. Balogh}

\affil{Department of Physics and Astronomy, University of Victoria, \\
 Victoria, B.C., V8W 3P6, Canada, \\
 email: Balogh@uvphys.phys.uvic.ca}

\author{R. G. Abraham}

\affil{Institute of Astronomy, University of Cambridge, Madingley Rd., \\
 Cambridge CB3 0HA, UK, \\
 email: abraham@ast.cam.ac.uk}

\and

\author{Tammy A. Smecker-Hane}

\affil{Department of Physics and Astronomy, University of California, 
 Irvine, CA 92797-4575, \\
 email: smecker@carina.ps.uci.edu}

\begin{abstract}
We discuss the galaxy population of the rich cluster MS1621.5+2640 at
z=0.4274, based on spectra and imaging in a field of size 9 arcmin by
23 arcmin ($\sim$2 by 5 h$^{-1}$ Mpc). The sample comprises 277
galaxies, of which 112 are cluster members, 7 are `near-members', and
47 are field galaxies in the redshift range 0.37$\le$z$\le$0.50. The
results are analyzed and compared with the z=0.2279 rich cluster
Abell~2390.  MS1621.5+2640 has a higher blue fraction, a younger
stellar population, and is a less evolved cluster. We do not find
strong evidence of significant excess star formation compared with the
field, although there is a small population of outlying near-members that is
unusually blue and that may be affected by the cluster. There is
a substantial population of red galaxies with significant H$\delta$
absorption, which are not easily explained by any simple form of
modeled star formation history. We detect two distinct
populations of cluster galaxies: those where star formation ceased some
time ago, and those with a gradual decrease over many Gyr. Our data
suggests that the cluster formed by accretion from the field, with
truncation of the star formation beginning at very large radii ($>$2
times the virial radius). The truncation process does not seem to be a
sharp one though, in that lower-luminosity early-type galaxies in the
inner core of the cluster are seen with significant H$\delta$
absorption, indicating some star formation within the last 1-2 Gyr. Some
combination of stripping of gas from the outer parts of the galaxy,
together with gradual exhaustion of the gas in the inner parts
would be consistent with our data.
\end{abstract}

\keywords{galaxies: clusters: individual (MS1621.5+2640) -- galaxies:
distances and redshifts -- galaxies: photometry -- galaxies: stellar
content}

\section{Introduction} \label{sec-intro}

Clusters at intermediate redshift allow us to study evolutionary
processes in cluster structure, cluster growth by infall, and the
evolution of cluster galaxies themselves, in morphology and stellar
populations. Study of a well selected set of clusters at different
redshifts provide snapshots of these processes at different epochs. The
Canadian Network for Observational Cosmology (CNOC) survey of selected
rich clusters (\cite{yee96a}, \cite{car96}) allows such studies and
comparisons over the approximate redshift range 0.2$\le$z$\le$0.6.

The first cluster studied in detail using the CNOC database was
Abell~2390 (\cite{abr96}, \cite{yee96b}), at redshift 0.2279. In those
papers, the cluster is seen to have well-defined  and smooth radial
gradients in galaxy color, morphology, and stellar population, showing
that the cluster forms by galaxies continuously accreting from the
field. This accretion process appears to involve the cessation of
star-formation, and passive evolution, creating a gradient in time
since last star formation with clustocentric distance. The overall blue
fraction of galaxies  in the field around Abell~2390 is higher than in
the zero redshift universe, both in the cluster and the nearby field,
consistent with general aging of all galaxy populations.

One controversy in this area has been how severely galaxies are affected
by accretion into a cluster: in particular, whether there is a
population of merging or interacting infalling galaxies, whether infall
disturbs the ISM enough to trigger a `starburst'\footnote{One immediate
cause for disagreement is in the definition of a `starburst'. Some
authors seem to regard an Sc galaxy as a `starburst'. We will try to
avoid some of this controversy by comparing our data with models with a
simple star formation history which we will specify. This will allow
others to decide whether any of the models are `starbursts' or not for
themselves. } in an isolated galaxy, whether gas stripping may occur
by the cluster IGM, and/or whether star-formation is halted on infall.
We noted that we see little evidence of a population with excess star formation in
Abell~2390, and advocated models of truncated star-formation to explain
the observed radial gradients. Recently, it was pointed out to us that
the issue of starbursting versus truncated star formation was
carefully discussed in \cite{new90}. Our conclusions here support and
expand on their work, as well as complementing that in \cite{abr96}.

Clearly it is of interest to see whether the same can be said of other
clusters, particularly those at even higher redshift, where the
fraction of blue star forming galaxies is known to be higher
(\cite{but84}).  This paper provides a comparison with Abell~2390 by
studying the rich cluster MS1621.5+2640 at z=0.4274. To allow easy
comparison, we tabulate some of the parameters of these two clusters in
Table~\ref{tbl-compare}. MS1621.5+2640 was chosen as the cluster in the
CNOC database most likely to violate the passive truncated star
formation model proposed for Abell~2390. We have data for a large
number of members, and it was noted when the observations were being
taken that line emission was common. As will be demonstrated, this
impression was largely driven by the fact that most galaxies (field and
cluster) at z=0.42 have higher star formation rates than those at
z=0.23. We find that MS1621.5+2640 does not have an unusually large
starforming population relative to the field.

A recent paper by \cite{bar96} studies the galaxy population
distributions in various observational parameter spaces, for the inner
regions of three z=0.31 clusters.  They conclude that some 30\% of the
cluster galaxies have undergone a secondary burst of star formation in
the previous 2 Gyr, and hence that infall causes a starburst. However,
the difference between a galaxy with truncation following a period of
rapid star formation, and one with simple truncation of star-formation
is small in their key diagram (H$\delta$ versus color, as discussed in
\cite{abr96} and \cite{new90}), since the `active' stage is very brief,
and the `post-starburst' stage (marked by strong H$\delta$ absorption)
is longer lived and indistinguishable from the `post-truncation' stage
of a truncation model. In fact, they do not clearly test and evaluate
the truncation model against a starburst model, and the number of
active rapidly star forming galaxies is very small and may not be
different from that of the field. \cite{pog96} use the same data, but
with independent models, coming to similar conclusions to \cite{bar96}.
We discuss this point in more detail in \S~\ref{sec-models}.

In \cite{bal97}, we took a different approach to the above - looking at
the [O~II] $\lambda$3727 Equivalent Width (EW) alone, but summing up
all the CNOC cluster data. Again we found no evidence for a
population with excess star formation relative to the field at the same redshift.
Cluster members were shown to have a lower star formation rate than the
field regardless of their color. The present detailed study of a single
well sampled cluster is meant to complement that more broad-brush
method.

Considerable interesting work on galaxy clusters at a range of
redshifts around 0.5 (overlapping the redshift of MS1621.5+2640) has
been published recently by the ``MORPHS'' collaboration (see
\cite{dre97} and references therein). They have used WFPC-2 data from
HST to morphologically classify the objects in these clusters, and
interpret these results as showing (amongst other results) that the
formation of ellipticals in clusters predates the formation of the rich
cluster, and also that S0 galaxies are generated in large numbers only
after the cluster virialization. It will be exciting to join our
results with those in the ``MORPHS'' papers, but we will not be doing
so in this paper. It is difficult to determine which of our current
sample of galaxies are morphologically S0, even using the concentration
index discussed in \S~\ref{sec-data}. The S0 class is a difficult one,
involving objects with a wide range of bulge to disk ratio, and
requires HST resolution to classify at these redshifts. We will be
attempting to make this link using HST imaging of MS1621.5+2640
scheduled to occur over the second half of 1998.

The outline of this paper is as follows: We describe briefly the data
and the line and morphology measurements in \S~\ref{sec-data}, and go
on to summarize the selection effects and biasses in the data in
\S~\ref{sec-bias}. Cluster membership criteria are explained in
\S~\ref{sec-sample}. The changes in the various measured galaxy
properties with location in the cluster are described in
\S~\ref{sec-dist}, and ROSAT HRI X-ray observations are presented in
\S~\ref{sec-xray}. Galaxy evolution models are described in
\S~\ref{sec-models}, and are used to interpret the observed gradients
in galaxy properties. Finally, we present our conclusions in
\S~\ref{sec-discuss}.

\section{Data and Measurements} \label{sec-data}

The data in this paper were obtained as part of the CNOC project. The
observational techniques and data reduction were the same as those for
Abell 2390 described by \cite{abr96}. The reader is referred to that
paper and also the CNOC data-reduction papers (\cite{yee96a},
\cite{yee96b}) for details of the experimental design. The data
catalogue for this cluster is given in full in \cite{ell97}.

The cluster MS1621.5+2640 was covered by central, northern and southern
9 arcmin by 7.9 arcmin fields, as it appears extended in the N-S
direction, giving a total field of 9 arcmin by 23.3 arcmin.  The
central field was observed with 3 spectroscopic masks, the Southern
with 2, and the Northern with 1. In all, 555 object spectra were
obtained, covering the observed frame wavelengths 4650-6300{\AA}
(3260-4410{\AA} in the cluster rest frame). From these, 277 reliable
redshifts were determined, of which 112 galaxies were classified as
cluster members. We discuss cluster membership in \S~\ref{sec-sample}
below.

\begin{figure}
\plotfiddle{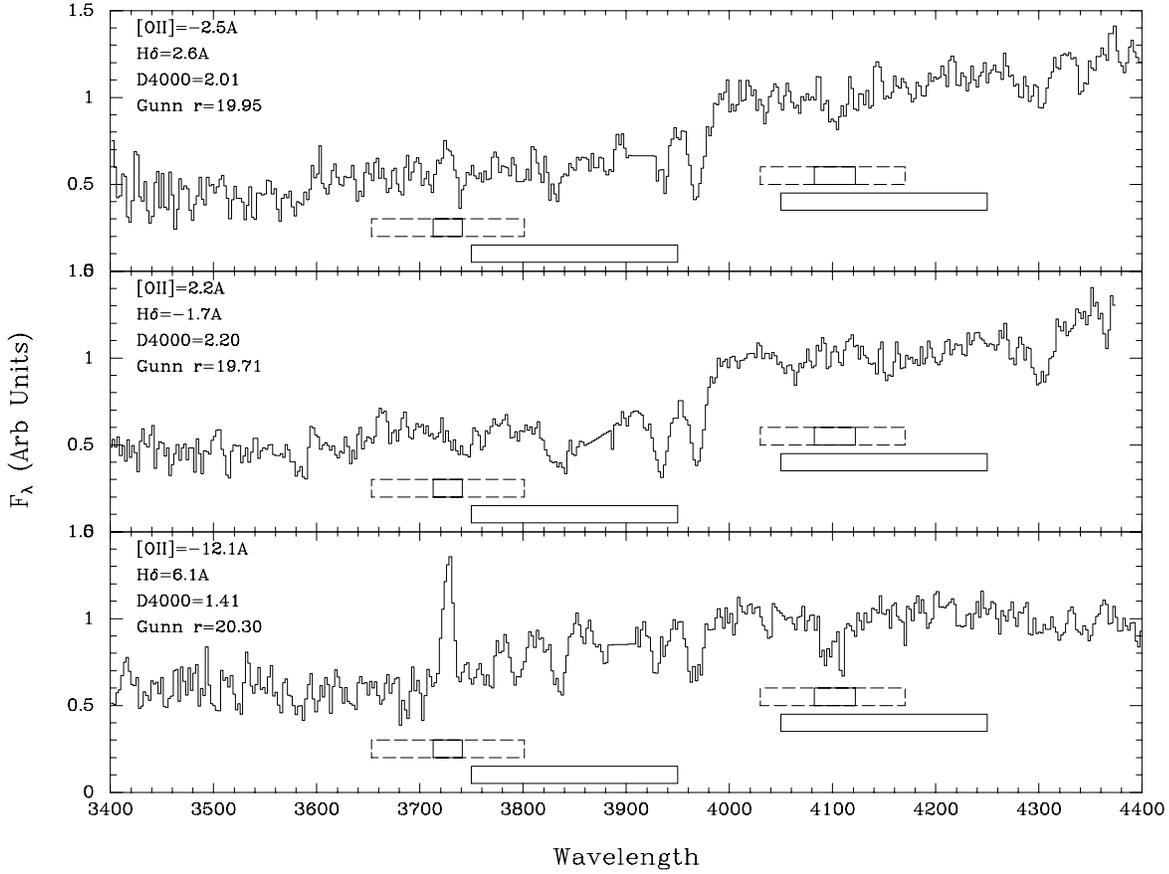}{4 in}{90degrees}{75}{75}{288pt}{-36pt}
\caption[fig1.eps]{Examples of line index definitions. The solid box
centered at 3727{\AA} is for [OII], the dashed boxes on either side
show its continuum regions. The solid box centered at 4101{\AA} is for
H$\delta$, with similar continuum bands. The D4000 index is defined as
the ratio of the fluxes in the two larger solid boxes centered around
3850{\AA} and 4150{\AA}. The three objects plotted (rest frame
wavelengths), from top to bottom are: (a) a red galaxy showing faint
signs of star formation including possible [OII] emission, and weak
H$\delta$ absorption. (b) a `normal' red galaxy. (c) A strongly star
forming galaxy. The line index values for each galaxy are shown in the
top left corners. The straight line segment just blue of the CaII
absorption is a residual from correction of the strong night sky line
at 5577{\AA} in the observed frame.
\label{fig1}}
\end{figure}

The fields were imaged with Gunn g and Cousins R filters (with the
latter calibrated in the Gunn r system), so that each galaxy has
measures of position, Gunn r magnitude, Gunn g-r color, morphology, and
several line indices from the spectra. Errors in the magnitudes and
colors are discussed in \cite{yee96a} and \cite{yee96b}.  The features
measured, and used in this paper, are D4000 (the 4000{\AA} break
strength), and the equivalent widths of [O~II] 3727{\AA}, and
H$\delta$. Indices and error estimates were calculated using the signal
and noise vectors along with the bandpass definitions described in
\cite{abr96} and \cite{bal97}. In
Figure~\ref{fig1}, we show some example spectra with the locations of
the index definitions plotted. These objects were chosen to illustrate
some of the range of indices seen. In the top spectrum, there is a red
galaxy with sizeable D4000, but also with possible weak [OII] in
emission and H$\delta$ in absorption. The middle spectrum shows how it
is possible to obtain a negative H$\delta$ index {\bf without} actually
having H$\delta$ in emission. If the continuum shape is such that the
continuum bands on either side of H$\delta$ are low then the index will
go negative. As will be explained later, the model spectra with which
we will be comparing are measured in exactly the same way, so this
unavoidable property of the index definitions should affect both data
and models in the same way, making comparison fair. The bottom spectrum
just shows a strongly star forming galaxy.

Updates to the measurement algorithm since the \cite{abr96} paper
include more robust error estimates, and also we have kept the two
measures of the H$\delta$ EW described in that paper separate, rather
than combining them with a break at an EW of 2{\AA}. We have only used
the `wide' (4082-4122{\AA}) definition in this paper. This simplifies
comparison with the models (see later), and does not seem to affect any
of the conclusions. We have studied all of these line indices to
investigate which are reliable measures and good tracers of stellar
population and age. As in \cite{abr96}, we have chosen to compare the
H$\delta$ and D4000 measures with galaxy evolution models, and use the
[O~II] as a measure of current star-formation activity.

We have also performed morphology measurements on the images, as
described in \cite{abr94}. In the original MOS images taken with the
spectra, the image quality is poor in all regions except the center of the
central field, so that reliable measures were possible only at small
radii. This is due to a combination of poor seeing ($>$1 arcsecond) and 
also the poor off axis image quality of the MOS spectrograph. Additional 
higher resolution images were subsequently obtained with the SIS 
spectrograph and tip-tilt correction
of the central field, and also used for morphology classification.
D. Schade is making independent morphological measures (private
communication), which will be used in a paper in preparation comparing
with the line indices and color for the whole CNOC sample. HST imaging
data for this cluster are scheduled to be taken, but are not yet
available.

An archive of all the CNOC data, including the spectroscopic indices
and the morphologies is being set up at the Canadian Astronomical Data
Center.\footnote{The CADC is part of the Dominion Astrophysical
Observatory in Victoria, B.C., which is part of the Herzberg Institute
of Astrophysics under the National Research Council of Canada.}

\section{Completeness and Biasses} \label{sec-bias}

The selection function for MS1621.5+2640 is plotted in
\cite{ell97}. This function covers all objects in the catalog - both
field and cluster members. One of the strengths of the CNOC sample is
that the field comparison sample can be drawn from objects observed in
exactly the same way as the cluster sample. This greatly reduces any
possible selection effects causing difference between the field and
cluster sample. As will be described in the next section, we also
restrict the redshift range of the field sample we use, in order to
further reduce the possibility of any such selection biases.

There are a number of reasons why a galaxy of a given magnitude may or
may not be in our spectroscopic sample. These are discussed in detail
in \cite{yee96a}, but to summarize the main points here, objects may be
too faint for a successful redshift measurement (magnitude selection
function), objects may be in a region that is too crowded for us to be
able to put slits on all objects using only 2-3 masks (geometric
selection), or objects may be lacking in strong features which are
needed for redshift measurement (color selection). In MS1621.5+2640, we
were unable to allocate enough slits to cover all the galaxies above
r=22, but have a high success rate for all such galaxies that we did
allocate slits to (see figure 3 in \cite{ell97}). Thus although we are
not complete, there should be no substantial biases in either the field
or member samples above this magnitude. This claim is supported by the
color selection functions shown in \cite{ell97} (figure 9). In
figures~\ref{fig12}-\ref{fig14} and figures~\ref{fig16}-\ref{fig17},
the small number of galaxies in the spectroscopic sample with
magnitudes fainter than r=22 are plotted with different symbols to
allow easy identification.

A second issue arises when we try to compare with the CNOC data on
Abell 2390 at z=0.2279. Are we probing to about the same lumnosity in
the two clusters, and hence is it fair to make comparisons of the two
samples without correction for this? The spectroscopic exposures for
MS1621.5+2640 were roughly 3 hours in duration, while those for Abell
2390 were 1 hour long. Additional factors affecting the relative depth
of the two samples are the CCD, grism and blocking filter efficiencies
at the two different observed wavelengths. This leads to color
selection effects in Abell 2390 starting to be significant at r=21
(\cite{yee96b}), roughly a magnitude brighter than in MS1621.5+2640.
Thanks to its lower redshift, it was easier to assign slits to a large
fraction of the galaxies in Abell 2390, and so the spectroscopic sample
is more complete, but comparing the points at which the magnitude
selection functions for the two cluster drop by a factor 2 gives that
the MS1621.5+2640 goes 1.3 magnitudes deeper in Gunn r than the Abell
2390 sample. The difference in distance modulus between the two
clusters is 1.5 magnitudes, and so we feel it is not neccessary to make
any correction for this.

Finally there is the question of comparisons between MS1621.5+2640 and
other clusters in the literature. The main comparison we will be
performing is with the data of \cite{cou87}, as analysed by
\cite{bar96}. Galaxies in their sample are drawn from a red (R$_{\rm
F}$=20) magnitude limited sample, which is close to the CNOC Gunn r.
They do not publish details of their selection function, but given that
their sample contains 112 galaxies from a projected radius from zero to
150 arcsec in three clusters at z=0.31, it is likely that they do not
contain as many low luminosity galaxies as we do. This may be the cause
of some of the differences between the conclusions in this paper and
that of \cite{bar96} discussed later.

\section{Cluster membership and field sample} \label{sec-sample}

\begin{figure}
\plotfiddle{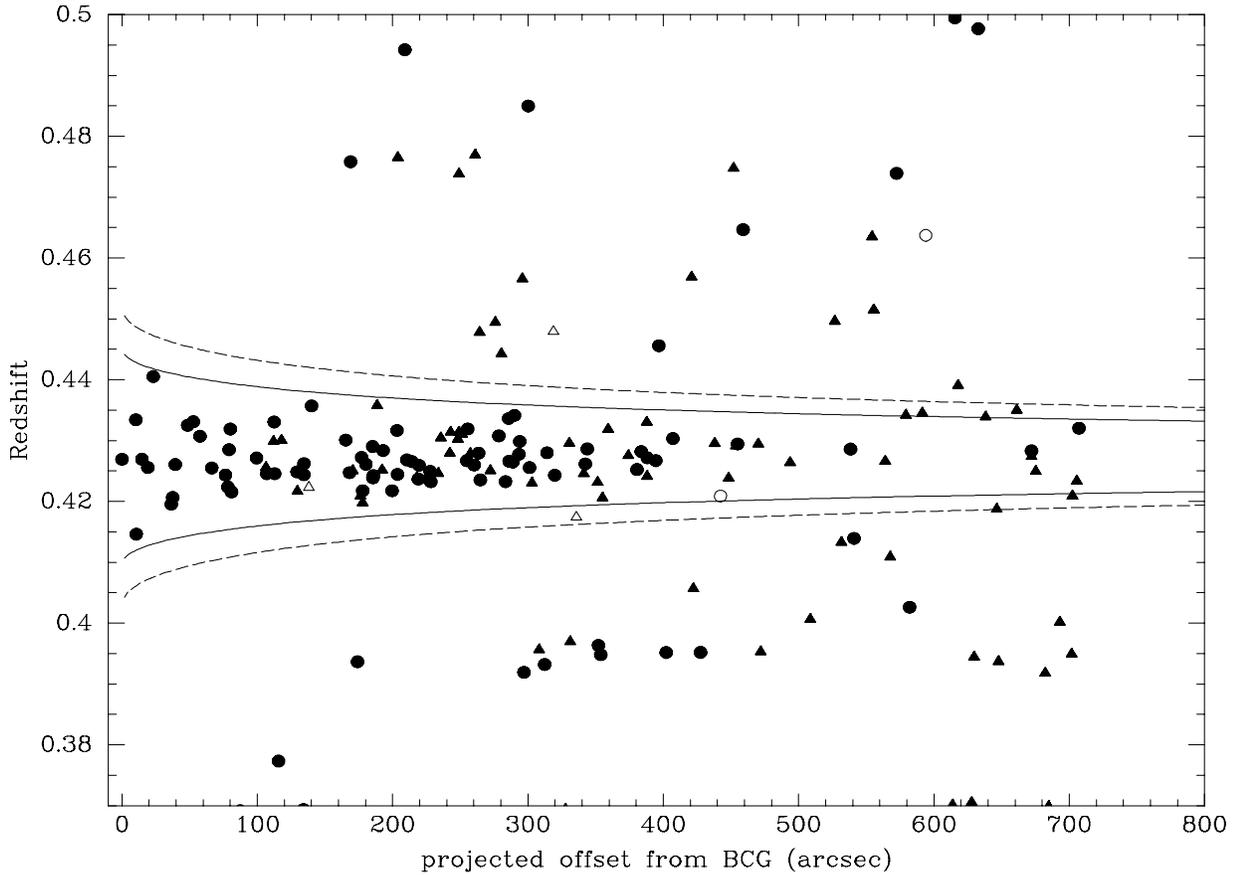}{4 in}{90degrees}{75}{75}{288pt}{-36pt}
\caption[fig2.eps]{Galaxy redshifts versus projected distance from
cluster center, separated into red (circles g-r$>$1.2) and blue
(triangles, g-r$<$1.2). The curves are 2.9 and 4$\sigma$ lines which
we adopt to define membership, as discussed in the text. Open symbols are
galaxies with poor S/N which are not used for the plots involving
spectroscopic measurements.
\label{fig2}}
\end{figure}

\begin{figure}
\plotfiddle{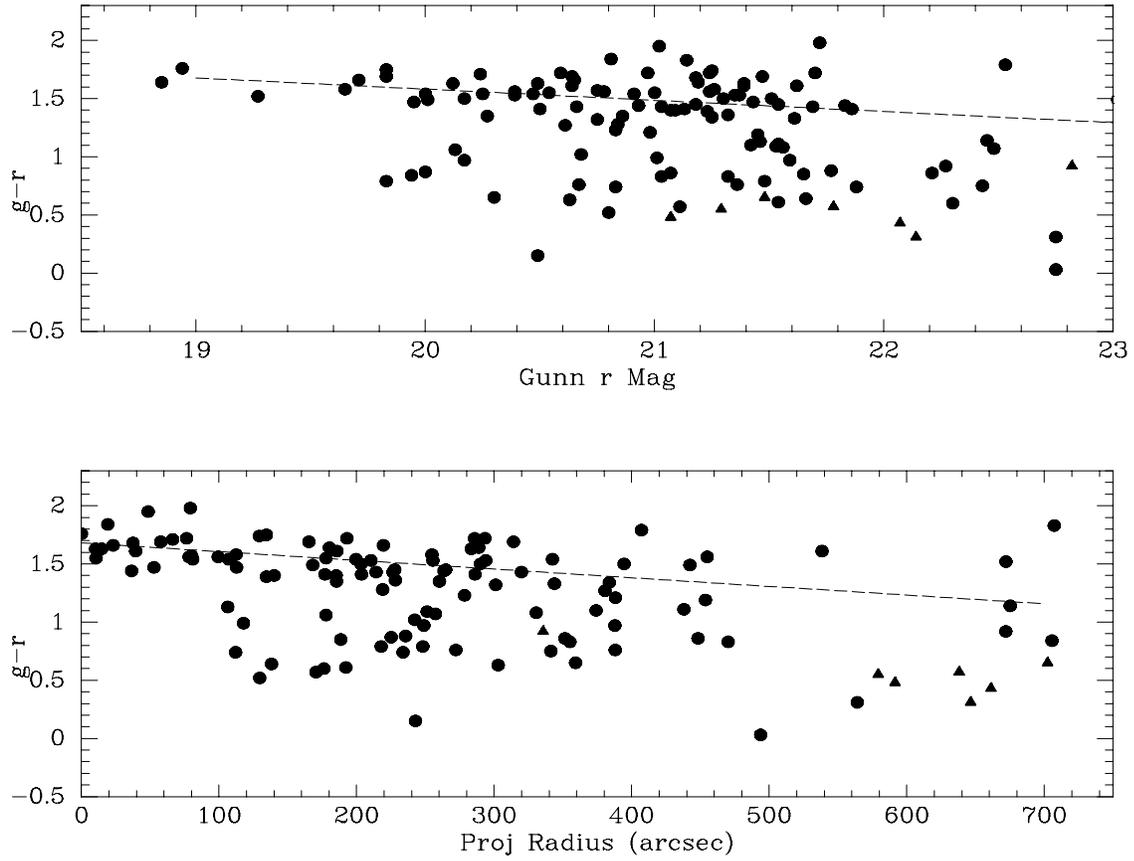}{4 in}{90degrees}{75}{75}{288pt}{-36pt}
\caption[fig3.eps]{Variation of cluster galaxy color with magnitude and
clustocentric projected distance. The lines are linear fits to them
with sigma-clipping to isolate the red population. The triangles are
the near-field galaxies. See \S~\ref{sec-bias} for a discussion of the unbiased 
magnitude range.
\label{fig3}}
\end{figure}

Figure~\ref{fig2} shows the redshift space near the cluster in our
data. Figure~\ref{fig3} shows that there is a well separated red
sequence of cluster members. This red sequence can be defined formally
by sigma clipping of the cluster galaxy colors, which converges to a
cut at  around g-r=1.2 (see below).  The distributions of red and blue
galaxies on the sky are very different (see Figure~\ref{fig5}), as in
Abell~2390, with the red galaxies showing a clearly defined `cusp' of
virialized members. The blue galaxies are not found in the cluster
center, and have a much less clear separation from the field at larger
radii.

The curves plotted in Figure~\ref{fig2} represent 2.9 and 4.0 times the
velocity dispersion of the cluster using the mass model of
\cite{car97}. The slightly odd value of 2.9$\sigma$ was chosen to
exclude the blue objects clumped near z=0.435 and projected radius of
600 arcsec. We take objects in the region between the 2.9 and 4
$\sigma$ curves as our `near-field' sample, and consider them
separately in our analysis. These (presumably unvirialized) objects are
bluer and have higher star formation rates than is typical even in the
field. Galaxies outside of the 4 $\sigma$ contour, but within the
redshift range 0.37 to 0.50 are used as a field comparison sample. This
is an arbitrary choice that gives roughly equal numbers of galaxies at
redshifts on either side of the cluster, with a small enough redshift
range to allow us to compare the important line features and make good
color k-corrections. The results are not sensitive to the field redshift
range used. We have a total of 112 cluster members with spectra, 47
galaxies lie in our field comparison sample, and 7 are near-field
galaxies. As will be discussed later, for all the analyses involving
spectral line indices, we further limit our sample using the error
estimates of the line indices. This reduces the above numbers to 110
members, 45 field and 6 near field galaxies (see \S~\ref{subsec-radial}
below).

Figure~\ref{fig3} shows the cluster members plotted on a
color-magnitude diagram, and a color-radius diagram. The red sequence
is fit  by sigma clipping of the sample.  The linear relations between
color and magnitude or radius are shown and have slopes of 0.096/mag
and $7.47\times10^{-4}$/arcsec (or 0.125/log(arcsec)). \cite{sta98}
demonstrate that the color-magnitude gradient is almost certainly
caused by a correlation between galaxy mass and metallicity. Because of
the small color change with magnitude (although somewhat larger than
that found in Abell~2390, which was 0.024/mag), we make no
color-magnitude correction in deriving the color-radius relation in
Fig~\ref{fig3}.  As in Abell~2390, there is a significant color change
with radius. For comparison, the slope in Abell~2390 is
0.079/log(arcsec). The same slope per kpc, scaled to z=0.4274, becomes
0.108/log(arcsec), consistent within the errors with that measured for
MS1621.5+2640. \cite{abr96} suggest that this color-radius relationship
is due to an age-radius relationship, produced by the gradual accretion
of field galaxies on to the cluster and their conversion into red, early
type galaxies.

\begin{figure}
\plotfiddle{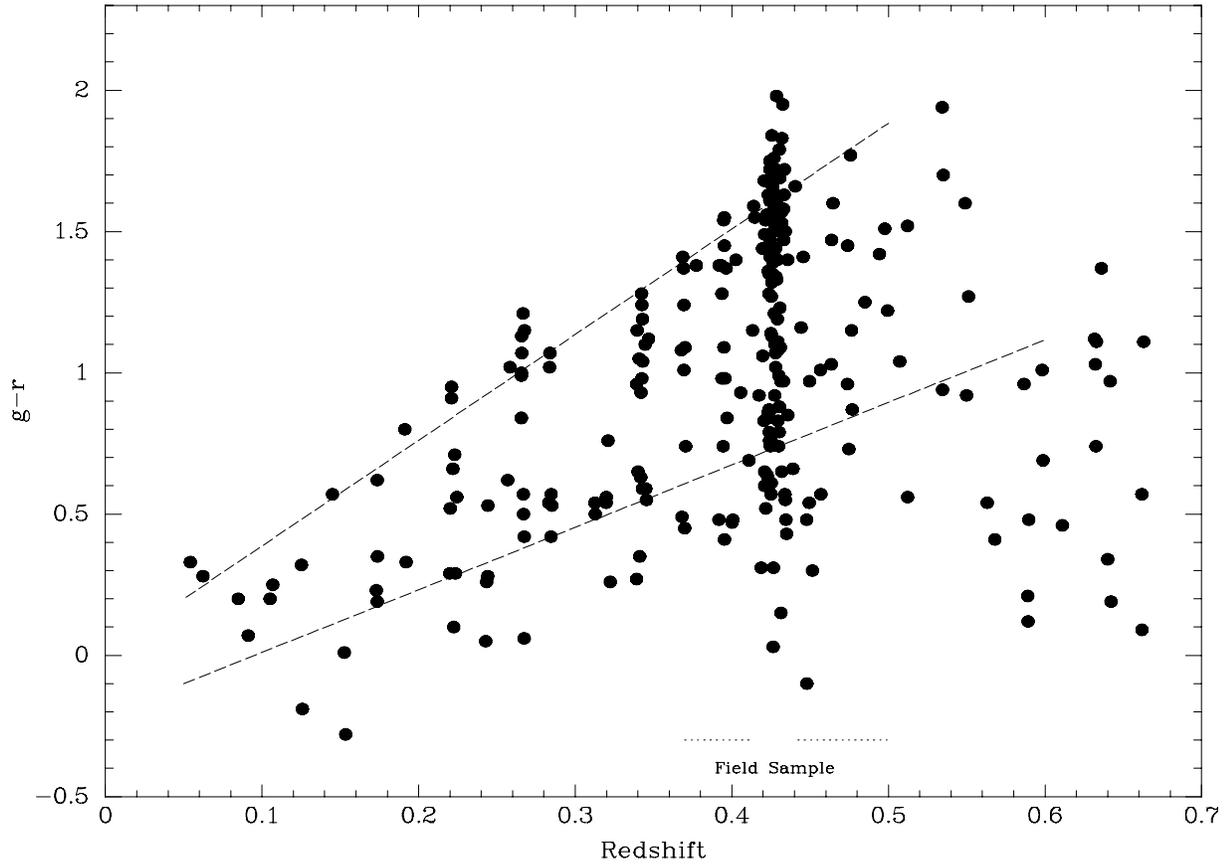}{4 in}{90degrees}{75}{75}{288pt}{-36pt}
\caption[fig4.eps]{Colors of all galaxies as a function of
redshift. The cluster and its red population stand out clearly. Other
structure can be seen at lower redshifts. The redshift range used for
the field galaxy sample is indicated. The lines illustrate the color
correction made to the field galaxies for comparison with the cluster
galaxies. 
\label{fig4}}
\end{figure}

Figure~\ref{fig4} shows the colors of all galaxies as a function of
redshift. To compare the field galaxies with the cluster members, we
adopt a color k-correction given by
$\Delta$(g-r)=(g-r+0.6)(z-0.4274)/(z+0.16), illustrated in the diagram
with two arbitrary observed color cases. This family of straight lines
is very close to the family of curves defined by different GISSEL
models (\cite{bru93}, and also \S~\ref{sec-models}) in the redshift
range covered by our field sample. As a check, we have verified that
the correction produces no color-redshift slope to within the
photometric errors in the data over the field sample range 0.37 to
0.50.

\section{Galaxy population distribution} \label{sec-dist}

\begin{figure}
\plotfiddle{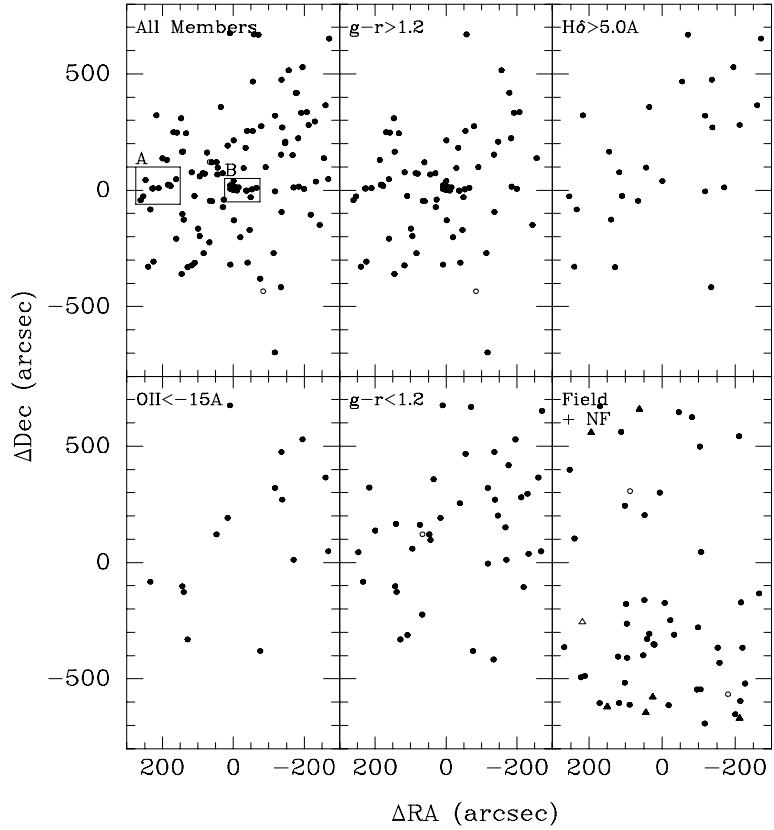}{4 in}{90degrees}{75}{75}{288pt}{-36pt}
\caption[fig5.eps]{Distribution of subsets of cluster galaxies and the
field sample.  The two subgroups discussed are outlined. Open symbols
are galaxies with poor S/N which are not used for the plots involving
spectroscopic measurements. Near field galaxies are plotted as
triangles. 
\label{fig5}}
\end{figure}

Figure~\ref{fig5} shows the distribution on the sky of galaxies of
different types. Galaxies plotted as open symbols have higher
spectroscopic errors, and are omitted from the plots based on those
measures, as discussed in \S~\ref{subsec-radial}.

In these plots we should note that the N-S distribution is uneven in
part due to the fact that only one North field mask was used for
spectra. This is obvious only in the field galaxy distribution:
conversely, the higher concentration of member galaxies in the North
may reflect a real asymmetry in the cluster population. The red
galaxies, separated as discussed above, dominate the core region within
100 arcsec.  The strong [O~II] emitters may be somewhat less centrally
concentrated than the strong H$\delta$ absorbers, consistent with
star-formation truncation with infall.

The member galaxy distribution, particularly in the red, suggests that
there may be a subgroup to the E of the core. We outline this group (A)
in Figure~\ref{fig5}, along with a box enclosing the cluster core (B).
Group A in fact contains the formal `Brightest Cluster Galaxy',
although this galaxy is 3 arcmin from the cluster center defined by the
X-ray centroid. The core group (B) coincides with the maximum in the
X-ray flux contours (see \S~\ref{sec-xray}). We discuss the co-added
spectra and group spectroscopic properties of the groups further in
\S~\ref{subsec-group}.

\subsection{Radial gradients} \label{subsec-radial}

In all analyses involving spectroscopic line indices, we have selected
the subset of the galaxies with H$\delta$ and [O~II] errors less than
10.0{\AA}. This just removes a few objects with extremely low S/N and
leaves us with nearly the full sample. We use this subset for all the
plots showing spectroscopic measures.  The subset contains 110 of the
112 cluster members, 45 of the 47 in the full field sample, and 6 of
the 7 near-field galaxies. As can be seen from the above numbers, this
leaves our sample largely unchanged, and all the discussion of
selection effects in \cite{ell97} and \S~\ref{sec-bias} should still be
appropriate. The resulting mean error in the [O~II] EW for cluster
members is 2.0{\AA}, and the median error is 1.7{\AA}.  The
corresponding numbers for H$\delta$ are 1.3 and 1.1{\AA}.  The D4000
errors are smaller proportional to the measurement (0.06 mean, 0.05
median), because this index involves averaging over many more pixels.

\begin{figure}
\plotfiddle{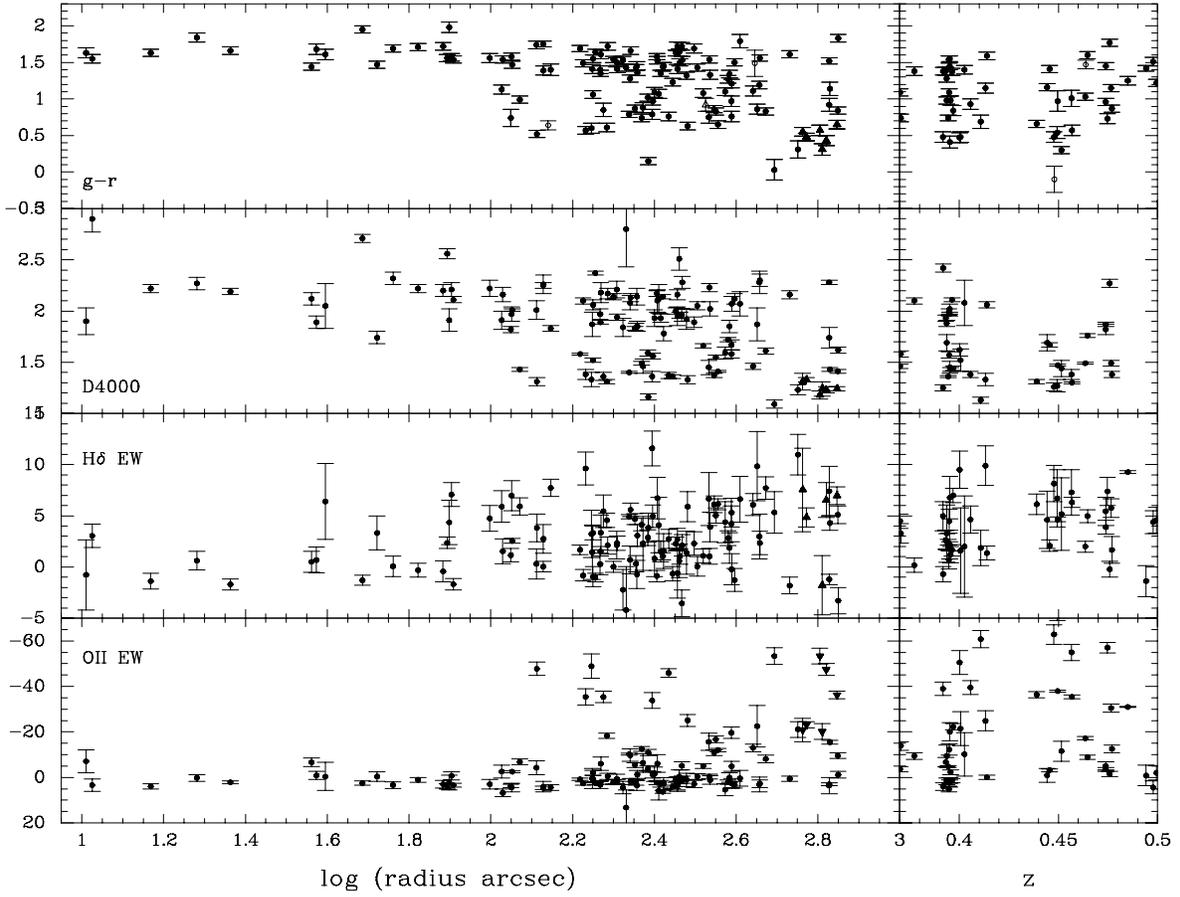}{4 in}{90degrees}{75}{75}{288pt}{-36pt}
\caption[fig6.eps]{Distribution of color and spectroscopic properties
with clustocentric projected distance. Triangles are the near-field
galaxies. The field samples are plotted against redshift for
comparison. For the top plot, this includes the full sample, the lower
three plots only show the subset with high quality spectroscopic
indices. Objects with large errors in line index are marked at open
symbols in the top panel.
\label{fig6}}
\end{figure}

Figure~\ref{fig6} shows the radial gradients of these measures with
error bars. We show the field galaxies distributed in redshift for
clarity of comparison.  This shows that there is a small foreground
group or cluster near redshift 0.395.

These figures show trends that are very similar to those found for
Abell~2390 by \cite{abr96}. The blue population grows with radius
within the cluster. This gradient is also reflected in the distribution
of the D4000 index. The [O~II] line emission is seen only in the outer
cluster (projected radius greater than 100 arcsec). For [O~II] emission
stronger than -15{\AA}, the cluster cannot be detected in redshift
space. This is consistent with the idea of truncated star-formation in
galaxies accreted from the field. If one excludes the `near-field'
objects (triangles), at no radius does the fraction of galaxies with
strong [O~II] exceed (or even match) that seen in the field. We also
note here that an [O~II] EW stronger than -15{\AA} is not unusual, even
at zero redshift. This is demonstrated, for example, by Figure~11 in
\cite{ken92}. The majority of spirals of class Sc for instance have
[O~II] this strong or stronger. The H$\delta$ absorption also shows a
decrease in the inner cluster, consistent with either pure truncation,
or a period of rapid star formation followed by truncation, since the
filling-in of Balmer absorption by emission during such a burst of star
formation is very short in duration.

The near-field galaxies may be remarkable, although the small sample
size makes any firm conclusions impossible. They appear to have extreme
values of all four indices in Figure~\ref{fig6}, even compared with the
field. A similar result is obtained for the near-field galaxies in
Abell~2390. This suggests that these galaxies, although supposedly not
yet virialized in the cluster, may have been affected by the cluster in
some way. It is perhaps appropriate at this point to mention that the
projected radius alone of the near field objects put them at a distance
of at least 2.3 h$^{-1}$ Mpc from the cluster core, more than twice the
R$_{200}$ `virialization' radius (\cite{car96}). Their velocity
difference, if due to Hubble flow, would add another 20 h$^{-1}$ Mpc in
quadrature to that distance.  What we are calling `near field' is
probably a long way from the cluster core as measured in virial radii.
A mechanism for increasing the star formation rate in galaxies at these
radii is unclear, but the phenomenon is worth looking for in other
clusters, to see if it can be established more definitely.

\begin{figure}
\plotfiddle{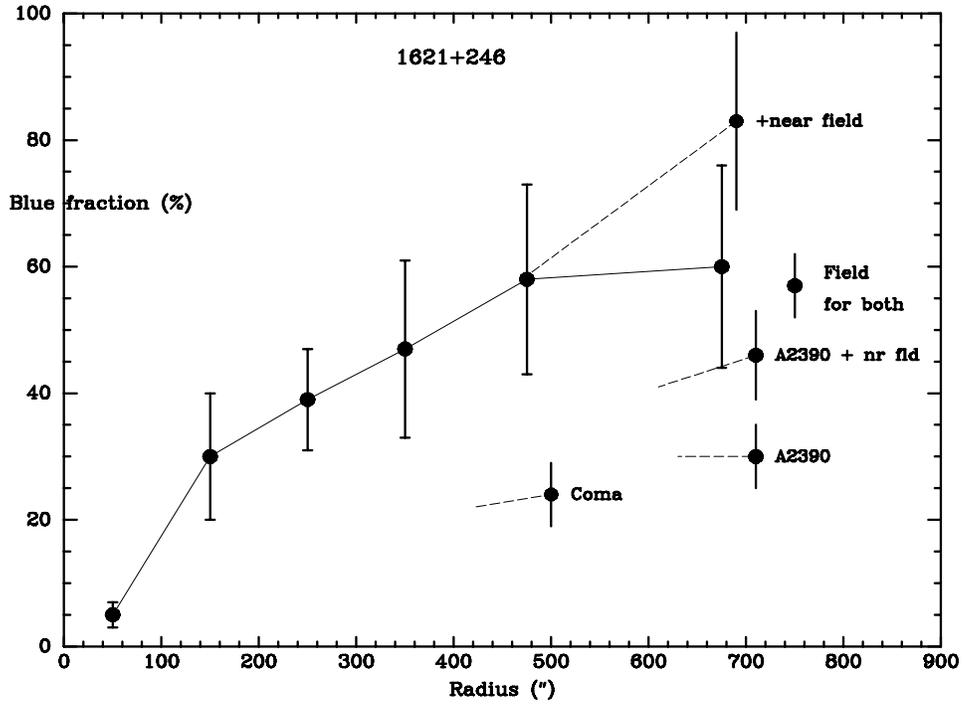}{4 in}{90degrees}{60}{60}{225pt}{0pt}
\caption[fig7.eps]{Blue fraction as defined by the $\sigma$-clipped
red population (see Figure~\ref{fig3}) as a function of projected
radius. The field value, and the value including the near-field galaxies,
are shown. The same quantities are shown for the outer parts of
Abell~2390 and Coma for comparison. (See \cite{abr96} for more detail
on the latter.) 
\label{fig7}}
\end{figure}

The strongest changes in the cluster population seen in
Figure~\ref{fig6} occur at a radius of about 100 arcsec.  Referring to
Table~\ref{tbl-compare}, we can see this corresponds to about 350
h$^{-1}$ kpc. This is a larger physical radius than that found for such
changes in Abell~2390 by about a factor 2.  In Figure~\ref{fig7} we
show the blue fraction of cluster members versus clustocentric radius.
The blue fraction uses a cut g-r=1.2, and is defined for both cluster
and field in the same way as the blue fraction shown for Abell~2390 in
\cite{abr96}. We show the outer cluster values for Abell~2390 and Coma
for comparison. The blue fraction in MS1621.5+2640 rises with radius
and reaches the field value, whereas in Abell~2390 it does not. This
difference could simply be due to the fact that the Abell~2390 data
does not go out as far in units of R$_{200}$ as the MS1621.5+2640
sample. We also show the higher value obtained when the near-field
galaxies are included, for both MS1621.5+2640 and Abell~2390. The
increase in blue fraction with redshift (\cite{but84}) is apparent, and
has been described for the full CNOC cluster sample by \cite{yee95}.

\subsection{Group properties} \label{subsec-group}

\begin{figure}
\plotfiddle{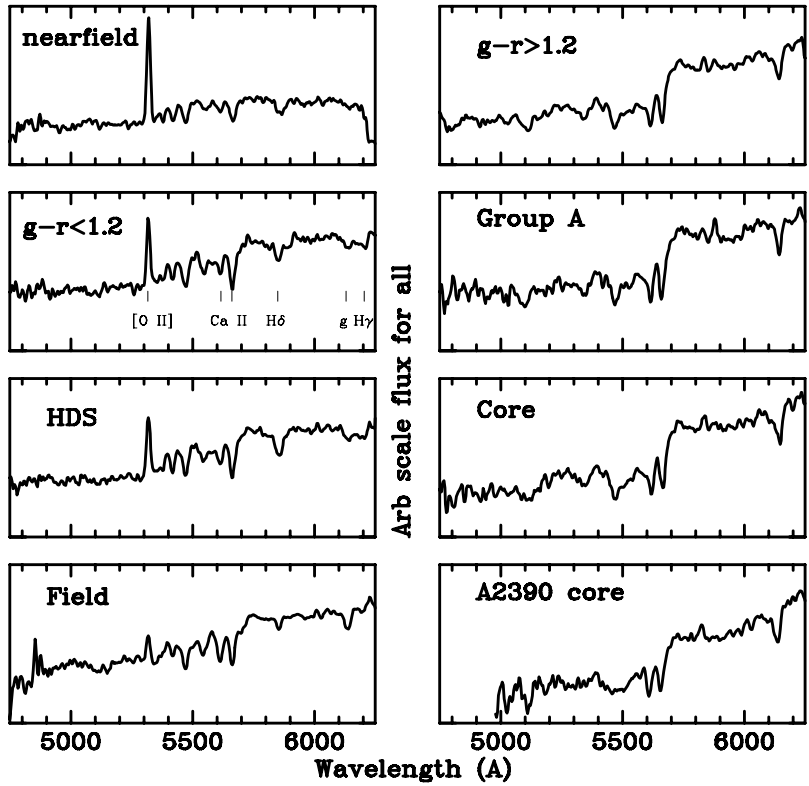}{4 in}{90degrees}{75}{75}{324pt}{-36pt}
\caption[fig8.eps]{Co-added spectra from subsets of galaxies, in the
observed wavelength frame. For comparison, the average of the central
galaxies of Abell~2390 are shown shifted to the redshift of
MS1621.5+2640. Fluxes are arbitrarily scaled for easy comparison.
\label{fig8}}
\end{figure}

Table~\ref{tbl-meanprops} shows the mean properties of various
subgroups. These quantities are derived from all the individual
galaxies involved with no cutoff or weights imposed as a function of
apparent magnitude apart from the one removing objects with errors in
their line indices greater than 10{\AA}. These groups are also
represented by co-added spectra, shown in Figure~\ref{fig8}. The
combined spectra are derived by adding the 10-15 brightest members of
each group, to maximize the S/N. The mean indices for the objects
summed to make the plot, and the groups as a whole, are not
significantly different, so the spectral plots can be considered
representative of the individual members of each group (although with
much higher S/N). Note that the values listed in
Table~\ref{tbl-meanprops} do not, in general, have Gaussian-shaped
distributions. There is no single number that measures their
spread, or the significance of the differences.  The main purpose of
Table~\ref{tbl-meanprops} is to allow typical values and trends to be
seen at a glance.

Figure~\ref{fig8} shows that the oldest population groups (Groups A and
B) in MS1621.5+2640 have spectra that differ from those in Abell~2390.
The difference is most noticeable in the CN feature (rest wavelength
~3875{\AA}, observed frame ~5500{\AA}), and is an indication that the
core Abell~2390 galaxies are older, as might be expected for a lower
redshift cluster.

The central group B forms an arc of galaxies that may correspond to an
X-ray feature (see \S~\ref{sec-xray}). These galaxies appear to be the
oldest, but also have an unusually large velocity dispersion. The
cluster is unusual in having the brightest member, cluster center, and
gravitational arcs all in different locations. It is likely that the
cluster is in a process of merging.

We also note that the group A galaxies have the properties and spectra
of old populations: thus, they appear to be an evolved subgroup that is
separate from the main cluster nucleus. They are not seen as a separate
X-ray flux peak, so are not associated with any detectable hot gas of
their own. The core subgroup, B, is slightly redder and more evolved (e.g. D4000).
The mean redshift of the core group (B) matches that of the cluster
members as a whole to better than 10 km/s. Group A has a mean redshift
offset from the cluster mean by 270 km/s in the cluster rest frame. We
suggest that these groups represent substructure, although we do not
discuss the cluster dynamics in this paper.

Table~\ref{tbl-meanprops} can also be used to determine whether there
are systematic differences in say galaxy magnitudes between the member
and field samples. Such differences could potentially invalidate the
comparison of spectroscopic indices. As can be seen, there are no
significant differences in mean luminosity between the blue member
galaxies and blue field galaxies or the red member galaxies and red
field galaxies.

\subsection{Galaxy morphology} \label{subsec-morph}

We have used the original MOS images for galaxy classification, as in
Abell~2390.  The image quality was not uniform across all fields, and,
in practice, only the center field was useable for classification. We
also have a higher resolution SIS image of the central 3 arcmin taken
in better conditions, which was used for classification of the central cluster
galaxies.

\begin{figure}
\plotfiddle{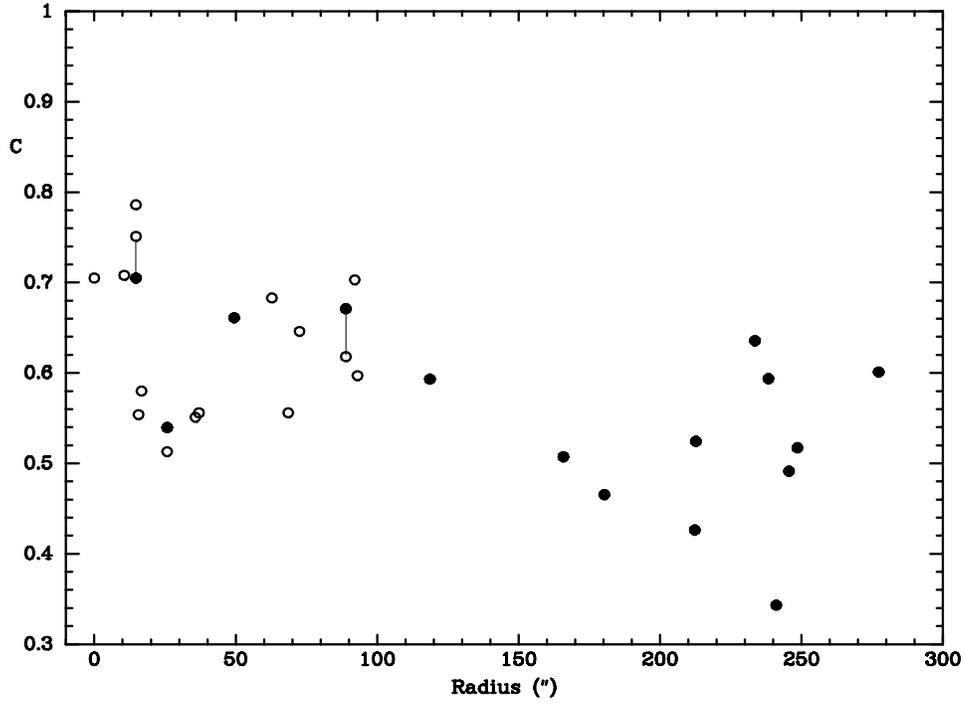}{4 in}{90degrees}{60}{60}{225pt}{0pt}
\caption[fig9.eps]{Galaxy central concentration for cluster members
with projected radius. The open points are from SIS image and the
filled points from MOS data. The zero-points are adjusted to minimize
the differences in values for the two galaxies in common, whose symbols
are connected. 
\label{fig9}}
\end{figure}

The concentration indices (\cite{abr94}) are shown in
Figure~\ref{fig9}. The measures for the two different resolution images
have different zero points as they are sampled differently and have
different depths. We have renormalized the SIS image points to the MOS
values based on two galaxies in common in the measures. There is an
overall trend to less centrally concentrated galaxies (measured by C),
with increasing radius, as seen in Abell~2390. There is a higher mean
value (0.64) in the inner 100 arcsec, and a lower value (0.5) in the
100-300 arcsec radius range. The small numbers and uncertain offset
between the two sets of measures make it difficult to draw more
detailed conclusions, but the overall trend matches that seen in
Abell~2390 in \cite{abr96}. The C values in Abell~2390 are higher and
fall more slowly as a function of projected radius, but detailed
comparison requires careful modeling of the effects of seeing and
redshift on the C index, which we felt was not justified by the quality
of the current imaging data for this cluster.

As mentioned in \S~\ref{sec-data}, independent morphological measures
are being made by D. Schade (private communication), and will be used
in a paper in preparation comparing with the line indices and color for
the whole CNOC sample. We are also obtaining HST images at several
radii for this cluster. Proper morphological statistics must await
these data.

\section{ROSAT HRI X-ray Emission} \label{sec-xray}

MS1621.5+2640 was observed with the ROSAT HRI July 28-30 1994 for a
total of 44093 seconds. Figure~\ref{fig10} shows a grayscale
representation of the HRI map with the locations of the galaxies with
measured redshifts in the CNOC sample overlaid. Circles indicate
cluster members, while plusses mark the positions of fore and
background galaxies with redshifts. `Near-field' galaxies are marked
as crosses. The HRI data was smoothed with a gaussian of width
$\sigma$=8 arcsec. Two very bright point sources are seen in the X-ray
which are not associated with any cluster galaxy. They are almost
certainly foreground stars or background AGN (both optical counterparts
are unresolved on the digital sky survey). We note that the extended
emission seen in this cluster in the EMSS catalog (\cite{gio90})
could well be almost all due to the nearest of the unassociated point
sources, given the comparatively poor spatial resolution of the
Einstein IPC. There is however diffuse amorphous emission seen in the
HRI image which is associated with the central group of galaxies in the
cluster. The S/N is insufficient to say much more than that the
emission is not strongly peaked on either the brightest galaxy of the
central group, or the galaxy surrounded by a giant arc. This amorphous
morphology contrasts strongly with the HRI data from A2390, presented
by \cite{pie96} which shows a much smoother profile. Other clusters
with similar X-ray morphologies to MS1621.5+2640 have been described in
the literature as examples of recent mergers (e.g. \cite{ulm82}). This
evidence for lack of equilibrium in MS1621.5+2640 is supported by the
following facts:
\begin{enumerate}
\item The brightest galaxy in the central regions is seen to have a
luminous close companion in images taken with higher spatial resolution
(unpublished CFHT data of S. Morris).
\item The bright arcs in this cluster are centered around the second brightest 
galaxy in the central group (\cite{lup92}, see also \cite{gio94}).
\item The genuinely brightest galaxy in the cluster lies 3 arcmin to the east 
of the cluster center (in group A of Figure~\ref{fig5}), and has no 
detectable X-ray emission associated with it.
\end{enumerate}
A detailed analysis of the X-ray gas and mass profiles derived for most
of the other CNOC clusters is given in \cite{lew98}. Unfortunately, the
HRI data quality for MS1621.5+2640 was sufficiently poor, and also the
evidence that the cluster X-ray gas was not in hydrostatic equilibrium
was so clear, that such analysis would not be appropriate for this cluster.

\section{Application of spectrum models} \label{sec-models}

As in \cite{abr96}, we have generated GISSEL models (\cite{bru93}) for
the redshift of the cluster (0.4274) and measured the same spectral
line and feature measurements as was obtained for the observed spectra.
Specific upgrades and changes since that paper include:
\begin{enumerate}
\item We are now using the Bruzual and Charlot 1996 models
(GISSEL96)\footnote{For experts, the majority of the models run used the 
spectral energy distribution files of the form bc96\_0pxxxx\_sp\_ssp\_kl96.ised, 
where xxxx is a number between 0004 and 1000 related to the metallicity 
(0200 being solar).}, copied from
gemini.tuc.noao.edu. These allow a range of metallicities to be
included.
\item These models are then shifted to the appropriate redshift and converted 
into IRAF format files, which are measured with exactly the same code as was used 
for the real data.
\item A correction has been computed for the fact that the new GISSEL96
models have lower spectral resolution than the observed data. Real CNOC
spectra were blurred out to match the model resolutions, and the
resulting changes in the indices measured. Details on this will be given in a
paper in preparation by Balogh et al., but roughly, this amounts to
adding 1.7{\AA} to the GISSEL model H$\delta$ values measured by the code.
\item Colors were measured using the IRAF `sbands' task. Magnitude zero
points were computed (also using this task but on appropriate A and G
type stars) to calibrate g-r.
\item A Salpeter IMF slope was used throughout. There are claims in the
literature that this IMF may in fact be universal (\cite{wys97}). We
will show one model where we imposed a `high' low mass cutoff at 2.5
solar masses, but otherwise all models include stars from 0.1 to 125
solar masses.
\item For Abell~2390, the models were reddened by E(B-V) of 0.075 (E(g-r)
of 0.084) using the \cite{bur82} dust maps. Higher resolution maps are
now available from \cite{sch98}, which give an E(B-V) of 0.11074 for
Abell~2390 and E(B-V) of 0.03143 for MS1621.5+2640. This latter value
was small enough that we have made no correction to these models for
galactic reddening.
\end{enumerate}
Two simple star formation histories were considered: 
\begin{itemize}
\item A 1 Gyr period of constant star formation starting at time 0, 
followed by complete cessation of star formation.
\item An exponentially decaying star formation rate  starting at time 
0, with a time constant of 4 Gyr (chosen to be significantly different 
from the previous model).
\end{itemize}
These are meant to roughly represent elliptical and late-type spiral
galaxies, respectively. The choices are arbitrary, but do produce
spectra that roughly match present day spirals and ellipticals at late times.
Obviously many far more complex star formation histories are possible,
and indeed may be more realistic, but it is intended that these two
simple cases will help illustrate the trends in the data.
Figure~\ref{fig11} shows how the two models evolve in the
color-H$\delta$ and color-D4000 planes for solar metallicity. Dots are
marked at 1,2,3...20 Gyrs. It can be seen that the two tracks combine
at early and late times (although note that the spiral model does not
reach the elliptical one until after 20 Gyrs for this model, which is
considerably longer than the age of the universe at a redshift of
0.4274). Between these two asymptotes, the spiral model spends a long
time with fairly blue colors, very slowly evolving towards the red. The
elliptical model spends a Gyr or so with elevated H$\delta$ absorption,
but rapidly evolves to red colors and low H$\delta$ (high D4000).

It should be noted that both of these models involve periods of time
when most people would describe them as `starbursts'. The elliptical
model in particular forms all its stars rapidly, and then spends a
period of 1-2 Gyrs looking like things generally called `poststarburst'
galaxies in the literature. Similarly, the spiral model in its first
few Gyrs spends time looking like a `starburst' or `poststarburst'
galaxy according to many definitions. As explained in a footnote
earlier, we would like to minimize the use of these terms, and leave
the reader to map from our model curves to their own terminology.

Models involving truncation of star formation in a spiral model
basically just jump across at constant H$\delta$ from the spiral track
to the elliptical track. Models including a burst of star formation
added on to a spiral model spend a very short time looping around to
very blue colors and low H$\delta$ (\cite{bar96}, \cite{pog96}), but
then move rapidly to parallel the elliptical track closely.
Without adding in a modified IMF, there are no models that fill in the
region of red colors and high H$\delta$. (We will discuss this further
in the next section).

We have also explored using the entire spectral model in comparison
with the observations, rather than measurements of only a few strong
individual features. This is most simply done by cross-correlation. The
peak height and width is a measure of the match:  in practice the lower
spectral resolution of the models makes this approach less sensitive
than it might be, and we do not achieve any greater sensitivity with
this approach compared using the most reliable individual features as
we have done.

\subsection{Modeling of the Observed Galaxy Population Gradients} \label{subsec-grad}

In Figure~\ref{fig12} we show the changes in the color-H$\delta$ plane
as a function of radial distance in the cluster. Moving from optimistic
to pessimistic, we note first that the GISSEL96 models with a simple
star formation history are a good match to the cluster data at radii of
$>$300 arcsec.  However, at smaller radii, the red galaxies all seem to
have H$\delta$ index too high for their color. Below we will explore
some possible ways to explain these objects, but will basically rule
out all the simple explanations that occur to us. This is not simply a
case of a few extreme galaxies, but seems to apply to {\bf all} the
galaxies in this color range. The radial trends mentioned earlier can
also be seen in these plots, with the `spiral' population rapidly
disappearing as one moves inwards in projected radius.

Figure~\ref{fig13} shows what seems to us to be astonishingly good
agreement between these simple models and the data for the color-D4000
plane. In particular, there is no evidence for large amounts of
reddening (which would be seen as a shift away from the model line
towards larger g-r), and there is a clear separation in the data
between objects matching the models with ongoing star formation, and
models where the star formation was truncated at some earlier time at
g-r$\sim$1.2.  Figure~\ref{fig14} shows that the excess of `H$\delta$
strong' objects can be seen (especially in the 100 to 300 arcsec radial
bin) even when D4000 is used rather than g-r color. This seems to
further exclude dust as a reasonable explanation, and also eliminated
problems in our magnitude calibration (either of the data or the
models) as the source of the mismatch. The data looks more scattered in
this plot, as the D4000 index has larger error bars than the g-r color,
and also the D4000 index is susceptible to systematic problems in the
spectroscopy not included in the formal random error bars, such as
occasional zero order or cosmic ray contamination, misplaced slits,
inaccurate extraction from the 2D spectral frames, etc.

An additional interesting fact about the red galaxies with significant
H$\delta$ absorption, is that they tend to be the lower luminosity
ones.  This is illustrated in Figure~\ref{fig15}. It is particularly
clear in the inner 100 arcsec panel, but in general the red galaxies
with H$\delta$ greater than 5{\AA} are all fainter than 20 in r. It
should also be noted though that our spectroscopic sample begins to be
biased in favour of strong line emitting galaxies at around Gunn r=22
(see \S~\ref{sec-bias}), as it is easier to derive redshifts for these
objects. Conversely, this will tend to remove objects from our sample
with red colors and weak absorption features fainter than r=22. 

Although the GISSEL96 models do not include line emission, we also show
the color-[O~II] plane in Figure~\ref{fig16} for completeness. It can
be seen that the red objects with excess H$\delta$ index over that in
the models basically show no [O~II] line emission at all. [O~II] is
seen exclusively in objects bluer than g-r=1.2. Comparison of the field
sample with the outer cluster region also re-emphasizes our earlier
point that at no radius is an excess of [O~II] emitters seen in the
cluster compared with the field, unless the near field objects are
included as members.

Finally in Figure~\ref{fig17} we show all the cluster members plotted
over the two simple models. This emphasizes the dramatic break between
objects with ongoing versus truncated star formation at g-r=1.2, and
also shows the general excess in the H$\delta$ index of all the red
galaxies relative to the models. The eye suggests that there may be two
populations of red galaxies - one with H$\delta$ less than 4{\AA}, and
a smaller, more extreme, group with H$\delta$ greater than 4{\AA}. The
star-forming and non-star-forming populations can be roughly divided at
a g-r=1.2 color, and so their spatial distributions can be read off
from Figure~\ref{fig5}. The blue galaxies are uniformly spread in
projected radius, and are not clumped towards the center, suggesting
they are not yet virialized. However, the RMS velocity of the blue
galaxies from Table~\ref{tbl-meanprops} is not significantly different
from that of the red galaxies. There is some evidence for reddening of
the red galaxies from the D4000 versus g-r plot by about 0.1-0.2 in
g-r, but this is not enough to explain the discrepancy in the top
plot.

We will now consider a few more extreme models to see if we can explain
the above mismatch. We show some of these in Figure~\ref{fig18}.
First, in the top left panel we test the effects of spectral resolution
on our measurements. For the comparison model, the 3520-7400{\AA}
region uses GISSEL models based on the spectra from stars in the
\cite{jac84} atlas. Apart from some significant differences for blue
objects with high H$\delta$ index, the higher spectral resolution
models match our corrected low resolution ones. Resolution effects
certainly do not move the models over into the region of red colors and
high H$\delta$ index. In the top right panel we show the effects of
reddening. Reddening of A$_{\rm v}$=1 would certainly allow the simple
models to explain the observations, but such massive amounts of
reddening, for a large fraction of the early-type population in the
cluster, is unreasonable, especially given the D4000 versus g-r plots
in Figure~\ref{fig13}. In the lower left panel we show the two most
extreme metallicity models in the GISSEL96 suite. Even 5 times solar
metallicity (a model which Bruzual and Charlot suggest should be
treated with caution in their documentation accompanying the GISSEL96
models) the g-r color does not move far enough to the red to match the
data.

Finally, prompted in part by \cite{cha93}, we show in the lower right
panel a model with the IMF truncated below 2.5 solar masses. This is
rather extreme truncation, but should illustrate the possibilities.
There have been suggestions in the literature for some time that the
IMF in `starburst' galaxies may be anomalous (see references in
\cite{cha93}), with diminished or no low mass stars. \cite{cha93} show
that in such cases, if the period of rapid star formation involves a
significant fraction of the galaxy's stars, then a galaxy can spend up
to a Gyr with unusually red colors and large D4000. As can be seen in
Figure~\ref{fig18}, the time spent with very red g-r colors is very
short for this particular model, and the time with simultaneously large
H$\delta$ index is even shorter. In this (again possibly extreme)
example, the model in fact spends a long time with unusually high
H$\delta$ and colors blueward of g-r=1 - a region unpopulated by
galaxies in our data. In summary, we do not feel that varying GISSEL96
model resolution, reddening, high metallicity, or bursts with
unusual IMFs will produce a good match to our observed red galaxy
distribution in H$\delta$ index versus g-r.

One caveat that should certainly be repeated here is that the H$\delta$
index we measure is just that - an index. As can be seen, it is
predicted to go negative at late ages, and indeed several red galaxies
do have such negative indices. This does not mean that they have
H$\delta$ in emission, but just that the continuum bands on either side
of the line are more depressed that the central band (see
Figure~\ref{fig1}). We have not yet investigated the details of what
molecules and ions dominate the absorption in this complex region, but
clearly this deserves further study.

\subsection{Comparison with Barger et al.} \label{subsec-barger}

As described earlier, \cite{bar96} present similar measurements for a
sample of 3 clusters at z=0.31. By combining the data from these
clusters, they end up with a sample of about the same size as ours (112
members total, compared to our 110 for MS1621.5+2640). The cluster
members for their sample range in projected radius from zero to 150
arcsec at z=0.31, and so generally correspond to our inner region in
Figures~\ref{fig12}-\ref{fig16}. The error bars plotted for H$\delta$
in \cite{cou87} are in general considerably larger than ours -
typically 2-4{\AA}, although \cite{bar96} claim the 1$\sigma$ errors
are in fact typically 1{\AA}. \cite{bar96} go on to model galaxy
evolution using an earlier version of the GISSEL models and also add in the
effects of line emission. They then quantitatively compare the numbers
of objects in specific regions of their diagrams (e.g.  their Figure~5)
with the expected numbers from the models, given the time spent in each
location. From this they conclude that around 30\% of the cluster
members had a secondary burst of star formation. We show their dividing
lines in Figure~\ref{fig19}, along with the numbers of galaxies in each
region in our data and in the \cite{bar96} data. There are a number of 
comments we would like to make about this comparison:
\begin{itemize}
\item We see considerably fewer `SB' (`starburst') class objects in our cluster.
\item We also see somewhat fewer `HDS' (H$\delta$ strong) objects. In
general, none of our current (simple) GISSEL96 models can match these, either as
objects with recently truncated star formation, or with an additional
burst of star formation. Although they do not explicitly plot the
models and the data together, we note that the GISSEL models shown in
\cite{bar96}, in our opinion, also have trouble reproducing the objects with red (B-V)
colors and high H$\delta$. The problem was also reported by
\cite{pog96}.
\item We do see a comparable number of objects in the `PSG'
(`Post-starburst' galaxy) region, but note that, given our error bars,
these points are all consistent with our two simple models that are
intended to match `normal' elliptical and spiral galaxies. At some
level this is semantics, in that our elliptical model does indeed
involve a `burst' of star formation, making up the whole galaxy.
However, the `PSG' points in our data set do not lie far (in terms of
the errors) from the `spiral' track with its smooth decay in star
formation. Objects in this region certainly are showing signs of recent
star formation - indicating a younger age, later formation time than
the majority of the cluster members.
\item Possibly due to our smaller error bars, it is easier to
(subjectively) identify populations of points that seem to go together.
If we modify the \cite{bar96} dividing lines, to use those shown in
Figure~\ref{fig19} as dotted lines, we then remove objects matching our
simple early-type galaxy model from the HDS category, and also remove
some objects consistent with our simple late-type galaxy model from the
SB and PSG category. Doing this generates the numbers in brackets in
the plots, and leaves rather few objects in our redefined `PSG' and
`HDS' catagories.
\item For comparison, we also plot the distribution and numbers of
galaxies in our field sample. Given the uncertainties from the
small sample size, the main differences are (a) a larger early-type
fraction in the cluster (no surprise at all), and (b) possibly an excess of
`HDS' galaxies in the cluster, although whether this ratio should be
taken relative to the whole sample, or alternatively just the other red
galaxies seem debatable to us.
\end{itemize}

\cite{pog96} perform an identical analysis to \cite{bar96}, using the
same data, but an independent galaxy modeling code. They reach similar
conclusions to \cite{bar96}, but also explicitly state that none of
their models (including a variety of bursts) can explain the objects
with B-R$>$2.3 and strong H$\delta$. Plotting their models over the
\cite{bar96} data confirms this. They also claim that their models can
not reproduce the large H$\delta$ values seen with simple truncation of
star formation and no burst. Obviously, we do not agree with this
conclusion. One possible reason for the disagreement might be that our
approach of using identical code on both the data and the models (with
a correction derived for the differing spectral resolution) is more
robust. However, it must also be admitted that we do not include the
effects of H$\delta$ infill by gaseous emission in our model plots,
which \cite{pog96} do.

Bringing in some IR data, \cite{bar96} and \cite{bar98} both look at
the evolution of the K-band luminosity function of the early-type
galaxies with redshift, and find no significant changes beyond passive
evolution between z=0 and 0.56.  This seems to support a model in which
there is no significant additional star formation in this population.

\subsection{Galaxy ages and metallicities} \label{subsec-ages}

In principle, comparison of the observed galaxy distributions with the
GISSEL96 models should allow rough determination of the galaxy ages. In
practice, as is well known, age and metallicity are inextricably
intertwined in most such comparisons. To illustrate this problem, we
show in Figure~\ref{fig20} the GISSEL96 D4000 and H$\delta$ indices
plotted against time. Overlaid on the model curves are the observed
distributions of the cluster members. For D4000, the well defined edge
around a value of 2.3 in turn leads to a fairly well defined age, if
the metallicity is known, or alternatively a well-defined metallicity
if the age is known. For solar metallicity, this value of D4000
corresponds to an age of 7.5 Gyr. For 2.5 times solar, the
corresponding age is 3 Gyr. Depending on one's preferred cosmology, one
might wish to invert this problem, and assume an age to derive typical
metallicities for the red galaxies\footnote{After this paper was
submitted a preprint from \cite{dok98} came out that uses the width of
the elliptical and S0 sequences in a color magnitude diagram of the
cluster MS1358.4+6245 (z=0.328, also in the CNOC sample) to set limits
on the star formation histories of these two populations. They use
analytical models for the evolution in color as a function of time, and
derive that star formation in the ellipticals ceased at z$\ge$0.6,
while the S0 population in the outer parts of the cluster experienced
star formation up to the epoch of observation.}.

One way to try to break this degeneracy would be by using some of the
other line indices (\cite{wor94}), but our other well studied index
(H$\delta$) has the problem that the entire red population seems to
have high H$\delta$ index relative to the models, making using the low
edge of the distribution (near -2 {\AA}) very questionable. We have not
looked in any detail at using other line indices from our spectra, or
alternatively folding in the g-r colors. This latter would obviously
also require estimation of the reddening in each galaxy.

\section{Conclusion} \label{sec-discuss}

We list below the main conclusions from the paper.
\begin{enumerate}
\item The blue population and general change in cluster membership
begins at a radius of 100 arcsec (or approximately 350 h$^{-1}$ kpc).
This is a larger linear radius than in Abell~2390 (approximately 150
h$^{-1}$ kpc).  Thus, the cluster core is considerably less compact in
MS1621.5+2640.
\item There is an evolved subgroup some 3 arcmin to the E containing
the true `BCG'.
\item The near-field galaxies may have enhanced star-formation. These
are at very large radii (greater than 2 h$^{-1}$ Mpc), and the
mechanism for this enhancement remains unclear.
\item A gradient is seen in the g-r colors of the red galaxies with
projected radius, which we interpret as a gradient in the time since
last star formation. This in turn fits well with models where the
cluster formed by gradual accretion of field objects, which in turn then
evolve (spectroscopically) into early-type galaxies.
\item The overall blue fraction is higher than at lower redshift, both
in the field and the cluster. The outer cluster blue fraction matches
that of the field, with the exception of the near-field (blue)
galaxies.
\item We find no signs of extensive star-formation in the cluster
compared with the field. Compared with the field, the average [O~II] is
weaker in the cluster, as is H$\delta$, and the galaxies are redder.
This is consistent with simple truncation of star-formation as galaxies are
accreted.
\item Detailed comparisons with GISSEL96 models show a good match in
general except for an odd excess in H$\delta$ index for all galaxies
redder than g-r=1.2. Various attempts to explain this excess
(resolutions differences, reddening, metallicity, truncated IMFs) were
not successful, but certainly deserve further study.
\item At this redshift, the models allow us to distinguish two galaxy
populations in the cluster: one best modeled with steadily decreasing
star-formation, similar to that seen in spirals in the field, and
another in which star-formation has terminated.
\end{enumerate}

To summarize further, and also indulge in some speculation, the above
data, combined with that on Abell 2390, seems to be consistent with a
fairly passive process of cluster formation. Galaxies are gradually
accreted from the field, with no good evidence for a large burst of
star formation as they are virialized into the cluster. The star
formation in the galaxies begins shutting down at very large radii
(well outside the virial radius), for reasons that remain unclear.
However, there is evidence for recent (within the last 1-2 Gyr) star
formation even in the inner 100 arcsec of the cluster. This star
formation is mostly seen in the lower-luminosity early-type galaxies in
the cluster core.  This suggests to us that the shut down process for
star formation could well be some combination of stripping of gas from
the outer parts of the galaxies as they fall in, followed by gradual
exhaustion of the gas in the inner parts. This would allow the remnants
of star formation to be still visible in galaxies seen in the core. One
hundred arcsec at this redshifts is 350 h$^{-1}$ kpc, which would take
0.4 Gyr to cross, going at the cluster velocity dispersion speed of 850
km/s. The fact that the inner galaxies with stronger H$\delta$ have
lower luminosities could either imply that they have had different
dynamical histories (i.e. the more luminous galaxies either formed
there or fell in earlier), or that the lower luminosity galaxies were
more gas rich, and hence more able to hang on to their star formation
in the presence of stripping. One should also remember the general
trend that more luminous galaxies even in the field tend to have formed
their stars earlier than the lower luminosity ones.

\acknowledgments

We would like to thank Amy Barger for helpful comments on an earlier
draft of this paper. The data used in this paper form part of the CNOC
study of intermediate-redshift clusters. We are grateful to all the
consortium members, and to the CFHT staff (particularly the Observing
Assistants Ken Barton, John Hamilton and Norman Purves) for their
contributions to this project.

\clearpage

\begin{deluxetable}{lrrrcccrrrr}
\small
\tablewidth{0pt}
\tablecolumns{3}
\tablecaption{Cluster Comparison of A2390 and MS1621.5+2640 \label{tbl-compare}}
\tablehead{\colhead{Property} & \colhead{A2390} & \colhead{MS1621.5+2640} }
\startdata
Redshift & 0.2279 & 0.4274 \nl
Scale\tablenotemark{a} (arcmin/h$^{-1}$ Mpc) & 6.897 & 4.615 \nl
Area Surveyed (arcmin) & $46\times7$ (E-W) & $9\times23$ (N-S) \nl
Area Surveyed (h$^{-1}$ Mpc) & $6.7\times1.0$ & $2.0\times5.0$ \nl
L$_{X}$ (h$^{-2}$ ergs/s, 0.3-3.5 KeV) & $4\times10^{44}$ & $5\times10^{44}$ \nl
Velocity Dispersion (km/s) & 1095 & 841 \nl
R$_{200}$\tablenotemark{b} (h$^{-1}$ Mpc) & 1.51 & 0.98 \nl
M$_{v}$\tablenotemark{b} (h$^{-1}$ M$_{\sun}$) & $2.6\times10^{15}$ & $0.98\times10^{15}$ \nl
\enddata
\tablenotetext{a}{q$_{0}$=0.1}
\tablenotetext{b}{see \cite{car96}}
\end{deluxetable}

\clearpage

\begin{deluxetable}{lrrrrrrrr}
\small
\tablewidth{0pt}
\tablecolumns{11}
\tablecaption{1621+246 mean group properties \label{tbl-meanprops}}
\tablehead{\colhead{Group} &\colhead{\#} &\colhead{g-r} &\colhead{D4000}
&\colhead{H$\delta$} &\colhead{[O~II]} &\colhead{R\tablenotemark{a}}
&\colhead{$\Delta$z} &\colhead{m$_R$}\\
&&&&\colhead{({\AA})} &\colhead{({\AA})} &\colhead{(")} }
\startdata
Blue (g-r$<$1.2)            &38 &0.82 &1.54 &5.1 &-15 & 320 &0.0038 &21.3 \nl
Red (g-r$>$1.2)             &72 &1.55 &2.09 &1.2 &  1 & 210 &0.0041 &20.8 \nl
\nl
HDS (H$\delta>$5)           &25 &0.99 &1.64 &6.9 &-13 & 319 &0.0041 &21.4 \nl
[O~II] ([O~II]$<$-15)       &15 &0.69 &1.40 &6.2 &-30 & 330 &0.0043 &21.8 \nl
\nl
Group A                     &10 &1.47 &2.00 &1.1 &  0 & 215 &0.0034 &20.8 \nl
Group B                     &12 &1.64 &2.19 &0.7 &  0 &  31 &0.0069 &20.7 \nl
\nl
Radius$<$100"               &19 &1.66 &2.22 &1.3 &  1 &  48 &0.0059 &20.8 \nl
Radius$<$100" g-r$<$1.2     & 0 & --- & --- &--- &--- & --- &  ---  & --- \nl
Radius 100"-300"            &59 &1.28 &1.87 &2.4 & -4 & 208 &0.0037 &20.9 \nl
Radius 100"-300" g-r$<$1.2  &20 &0.81 &1.53 &4.5 &-16 & 202 &0.0042 &21.1 \nl
Radius$>$300"               &32 &1.12 &1.75 &3.7 & -7 & 439 &0.0029 &21.3 \nl
Radius$>$300" g-r$<$1.2     &18 &0.83 &1.54 &5.8 &-13 & 451 &0.0034 &21.5 \nl
\nl
Near field                  & 6 &0.49 &1.25 &7.3 &-34 & 636 &0.0076 &21.6 \nl
\nl
Field                       &45 &1.06 &1.62 &4.0 &-18 &(436)&   --- &21.1 \nl
Field g-r$<$1.2             &27 &0.80 &1.42 &5.1 &-28 &(470)&   --- &21.4 \nl
Field g-r$>$1.2             &18 &1.43 &2.01 &2.3 & -3 &(384)&   --- &20.7 \nl
\enddata
\tablenotetext{a}{Projected distance in arcsec from cluster center}
\end{deluxetable}

\clearpage

\clearpage

\clearpage

\clearpage

\clearpage

\clearpage

\clearpage

\clearpage

\clearpage

\clearpage

\begin{figure}
\caption[fig10.eps]{(Removed due to problems with LANL policy on file
sizes. Can be obtained from: http://www.hia.nrc.ca/STAFF/slm/ms1621)
ROSAT HRI map with the locations of the galaxies with measured
redshifts in the CNOC sample overlaid. Circles indicate cluster
members, while plusses mark the positions of fore and background
galaxies with redshifts. `Near-field' galaxies are marked as crosses.
Black on the grayscale corresponds to an HRI detection rate of roughly
$3\times10^{-6}$ counts s$^{-1}$ arcsec$^{-2}$. The region plotted is
slightly larger than the spectroscopically surveyed field of 9 arcmin
by 23.3 arcmin.
\label{fig10}}
\end{figure}

\clearpage

\begin{figure}
\plotfiddle{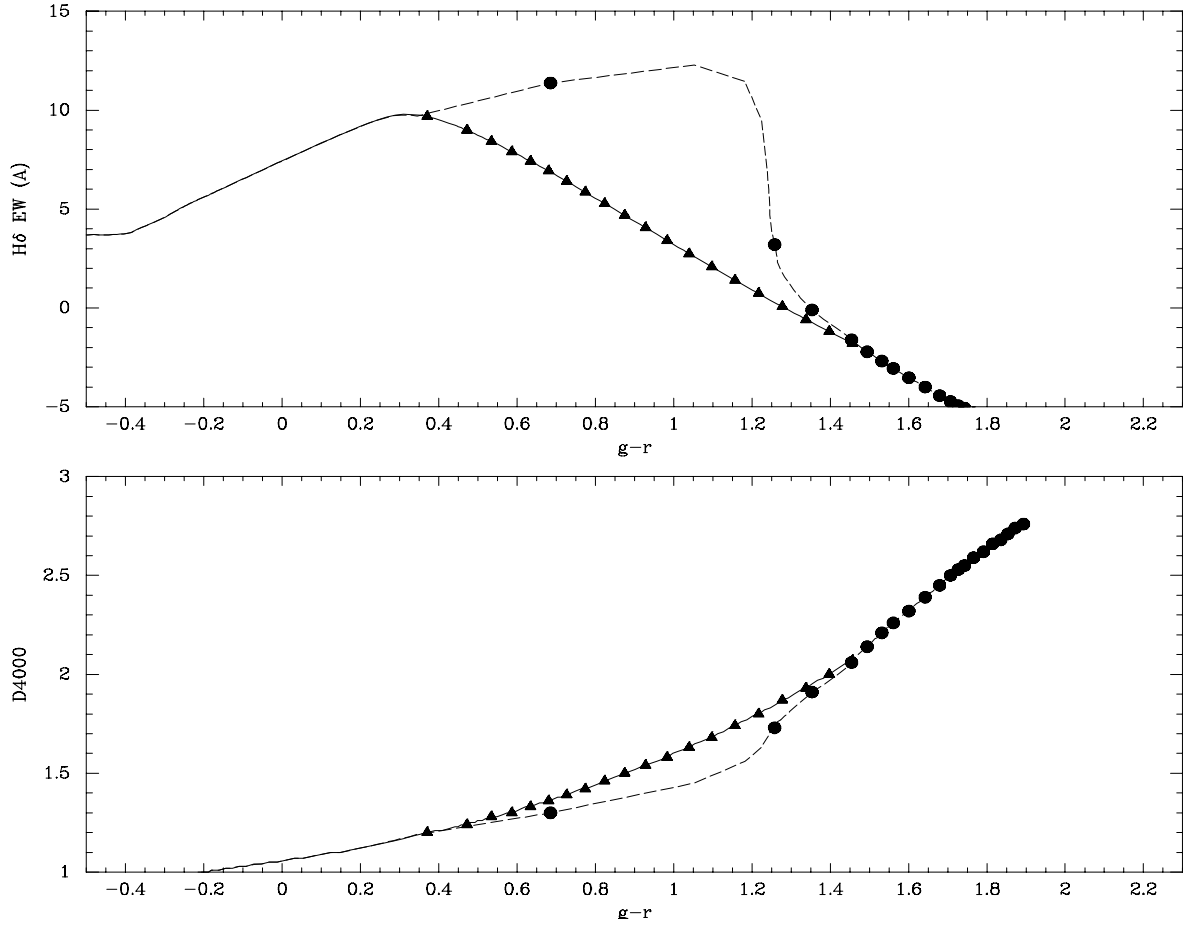}{4 in}{90degrees}{75}{75}{288pt}{-36pt}
\caption[fig11.eps]{GISSEL96 models (\cite{bru93}) with Salpeter IMF and
solar metallicity, for redshift 0.4274. The dashed line shows a track
for 1 Gyr of continuous star formation followed by complete cessation.
The solid line shows exponentially decreasing star-formation with a
time constant 4 Gyr. Symbols indicate ages at 1 Gyr intervals from 1 to 
20 Gyrs.
\label{fig11}}
\end{figure}

\clearpage

\begin{figure}
\plotfiddle{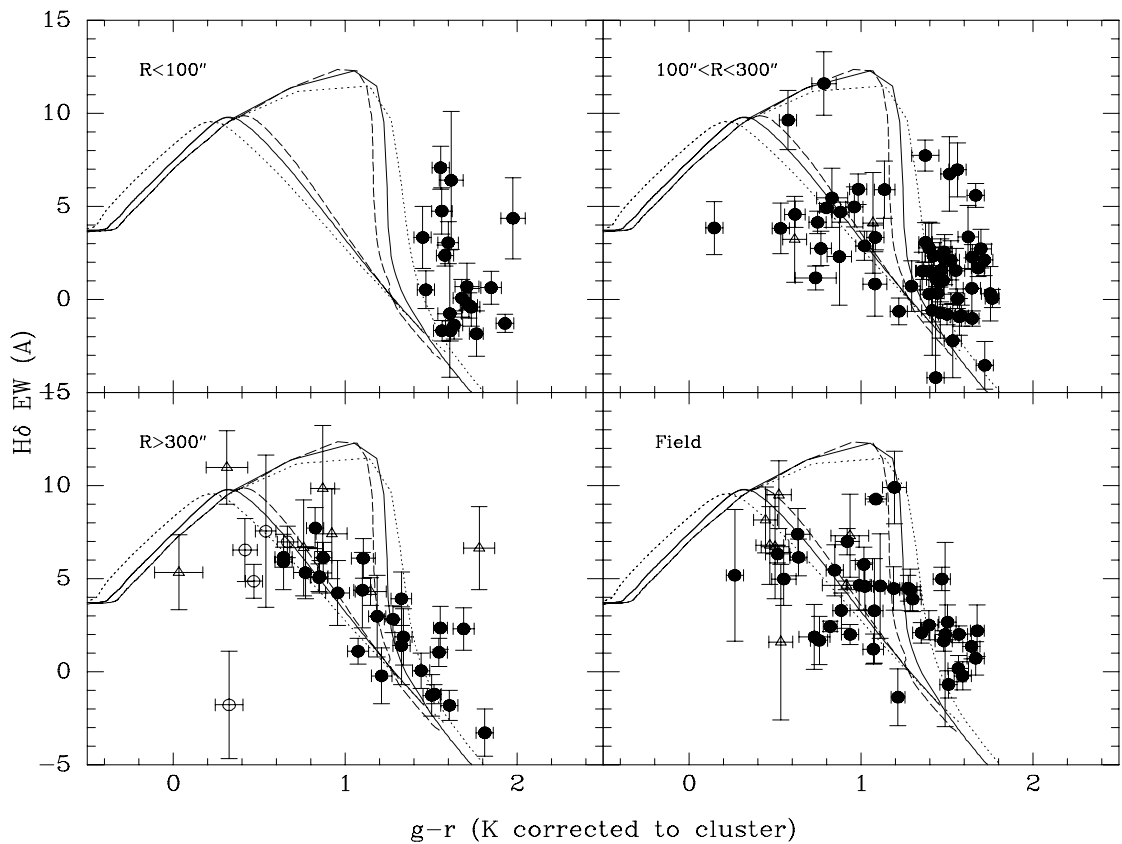}{4 in}{90degrees}{75}{75}{288pt}{-36pt}
\caption[fig12.eps]{H$\delta$ versus g-r color for 3 projected radius
bins in the cluster and for the field (k-corrected to the cluster
redshift). Open triangles are member galaxies with r magnitude fainter
than 22.0 (see \S~\ref{sec-bias}), open circles are the near-field
galaxies.  The lines sketch the models illustrated more clearly in
Figure~\ref{fig11}. The solid line shows solar metallicity, the dashed
line is 0.4 times solar, and the dotted line is 2.5 times solar.
\label{fig12}}
\end{figure}

\clearpage

\begin{figure}
\plotfiddle{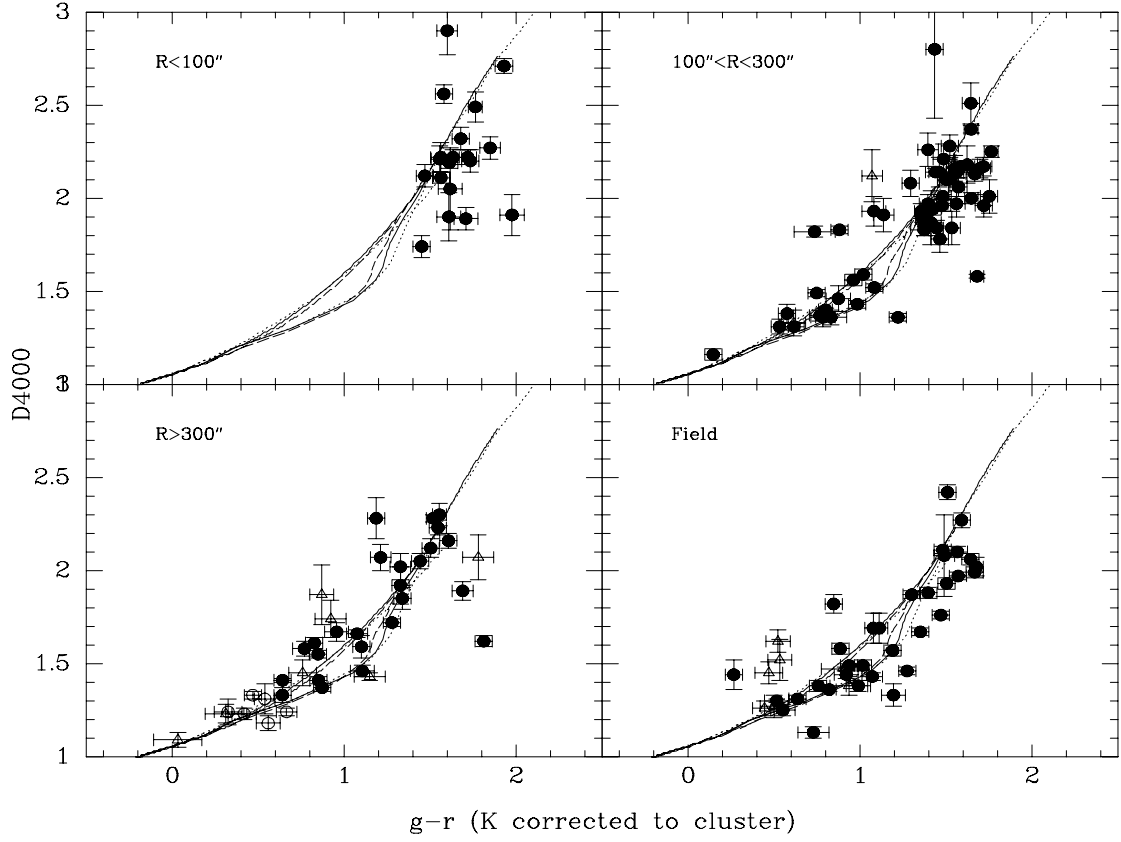}{4 in}{90degrees}{75}{75}{288pt}{-36pt}
\caption[fig13]{As Figure~\ref{fig12} but for D4000. 
\label{fig13}}
\end{figure}

\clearpage

\begin{figure}
\plotfiddle{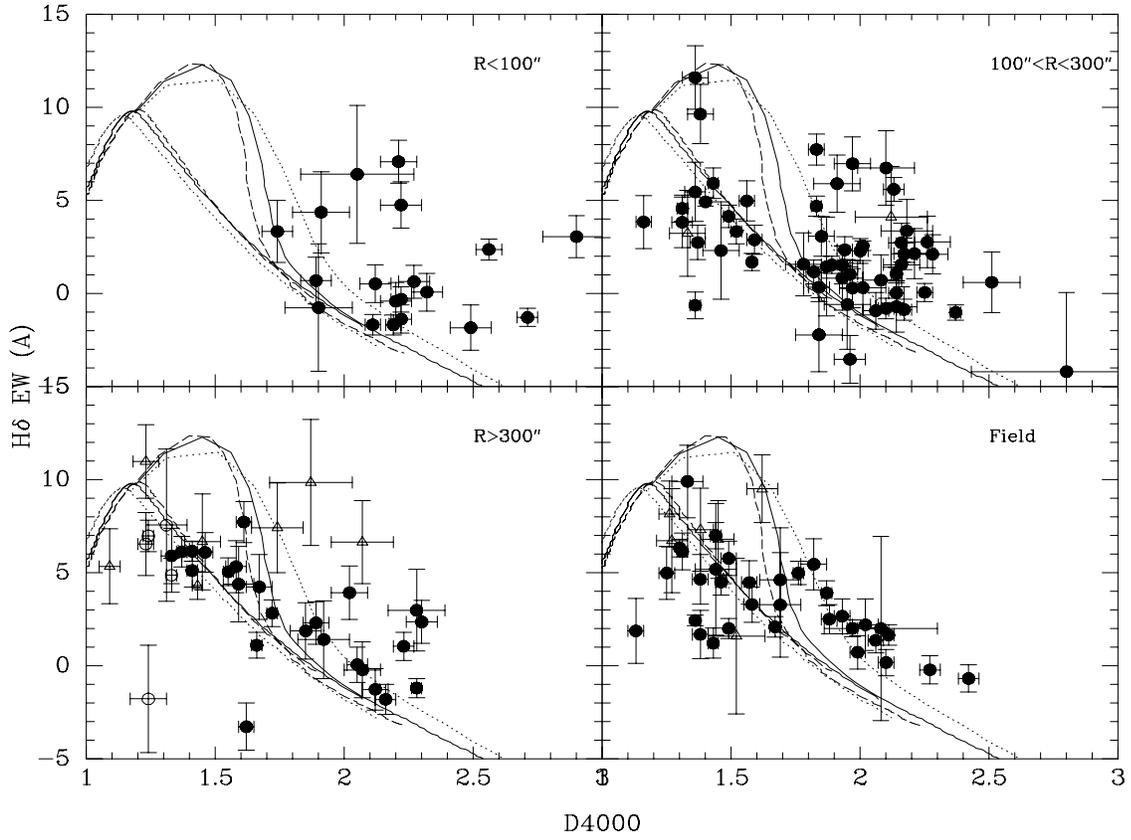}{4 in}{90degrees}{75}{75}{288pt}{-36pt}
\caption[fig14]{As Figure~\ref{fig12} but plotting the two
spectroscopic measures to avoid reddening uncertainty. 
\label{fig14}}
\end{figure}

\clearpage

\begin{figure}
\plotfiddle{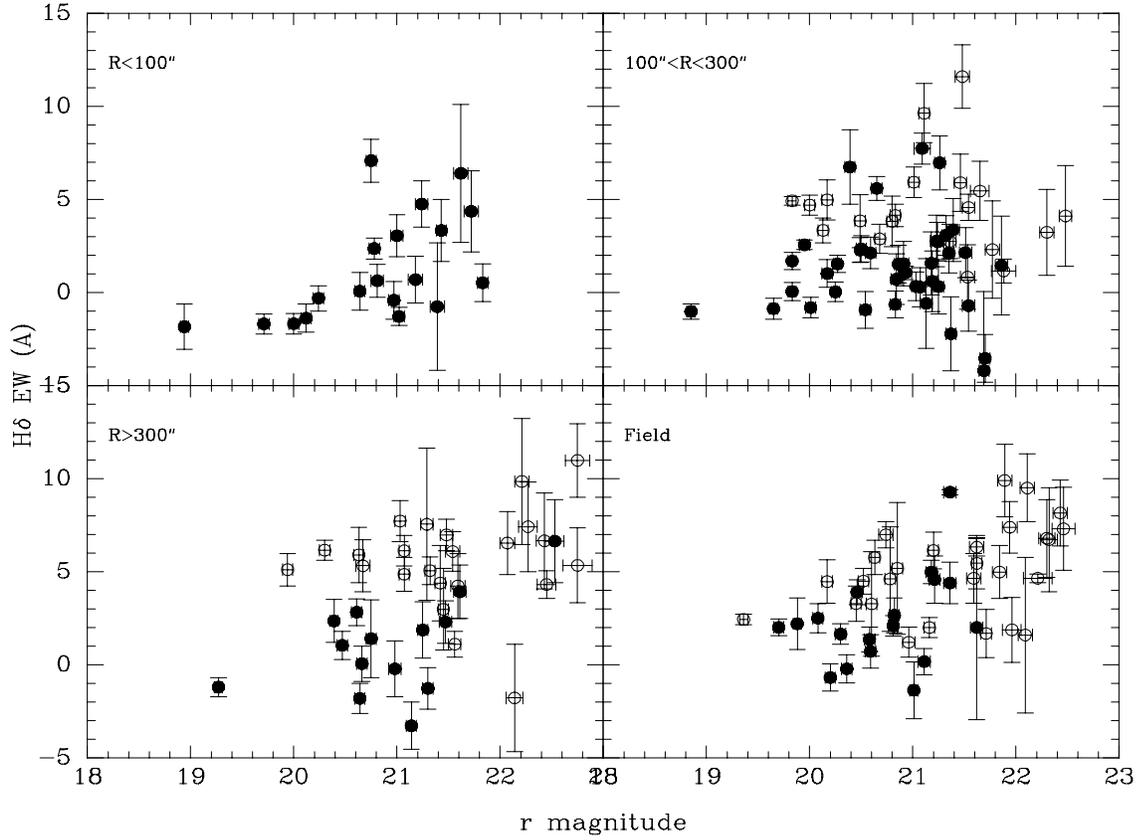}{4 in}{90degrees}{75}{75}{288pt}{-36pt}
\caption[fig15]{H$\delta$ versus apparent r magnitude. As all the
objects in the cluster (and indeed the field sample) are at roughly the
same redshift, this is proportional to luminosity. Filled symbols are
galaxies with g-r color $\ge$1.2 (red), while open symbols are galaxies with
g-r color $<$1.2 (blue). See \S~\ref{sec-bias} for a discussion of the unbiased 
magnitude range.
\label{fig15}}
\end{figure}

\clearpage

\begin{figure}
\plotfiddle{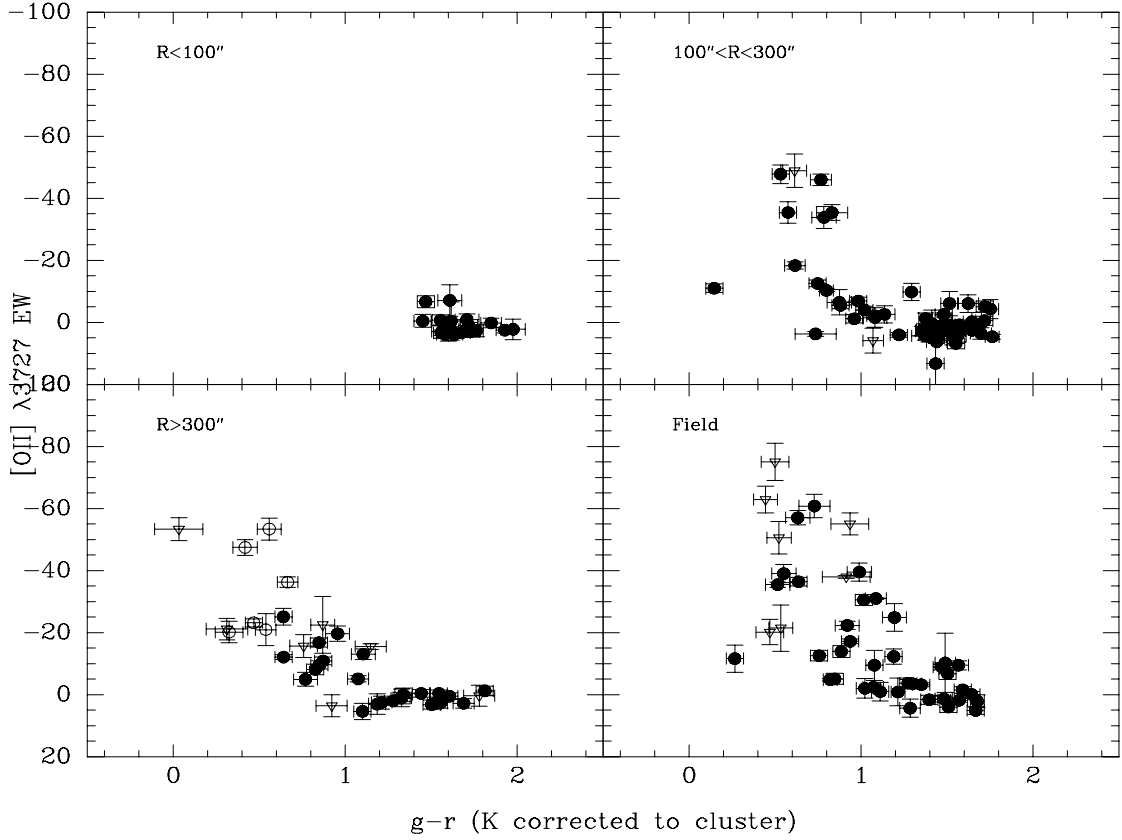}{4 in}{90degrees}{75}{75}{288pt}{-36pt}
\caption[fig16.eps]{As Figure~\ref{fig12} for [O~II] emission.
The  GISSEL models (\cite{bru93}) do not include emission lines, 
and hence no model lines are plotted. 
\label{fig16}}
\end{figure}

\clearpage

\begin{figure}
\plotfiddle{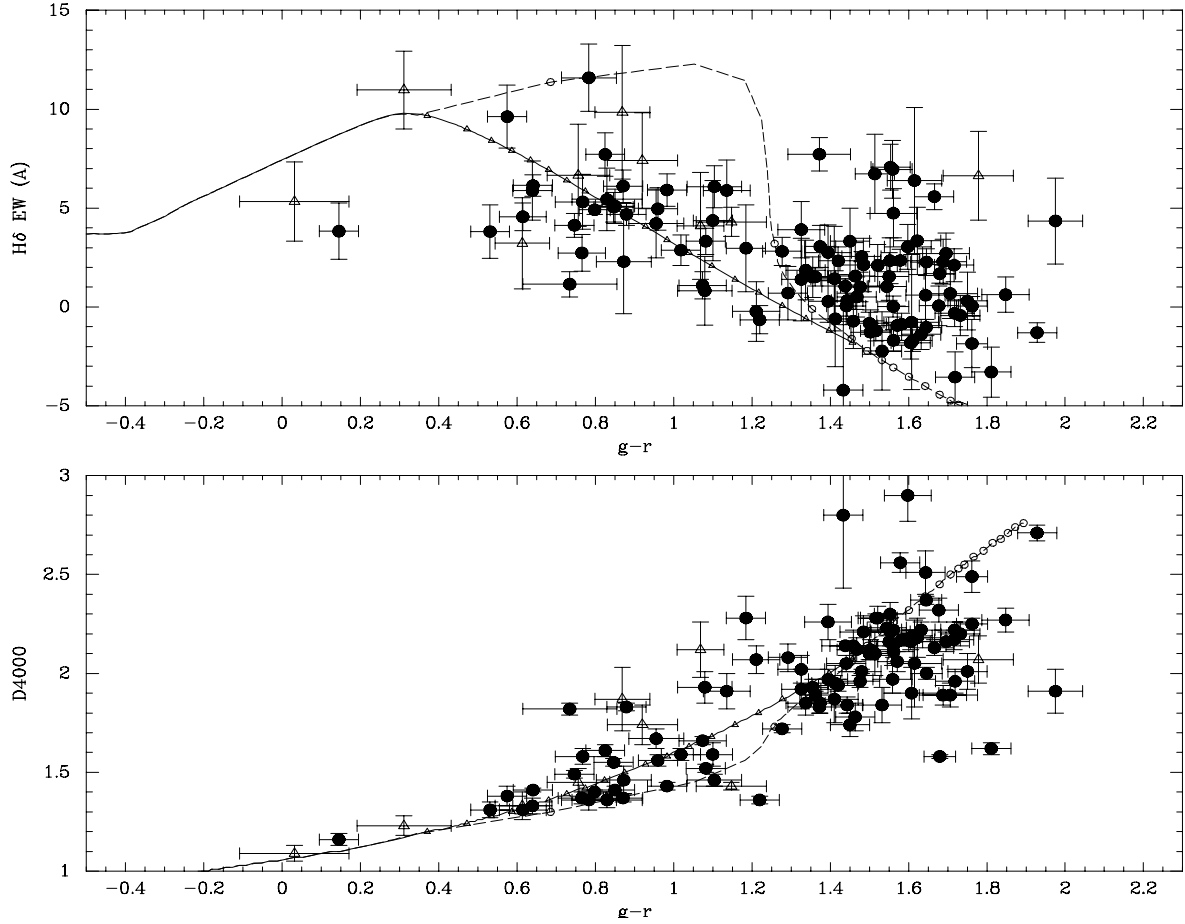}{4 in}{90degrees}{75}{75}{288pt}{-36pt}
\caption[fig17]{Measured quantities for all cluster members (i.e.
summing over all radii) compared with solar metallicity models from
Figure~\ref{fig11}. Open triangles are member galaxies with r magnitude
fainter than 22.0 (see \S~\ref{sec-bias}).
\label{fig17}}
\end{figure}

\clearpage

\begin{figure} 
\plotfiddle{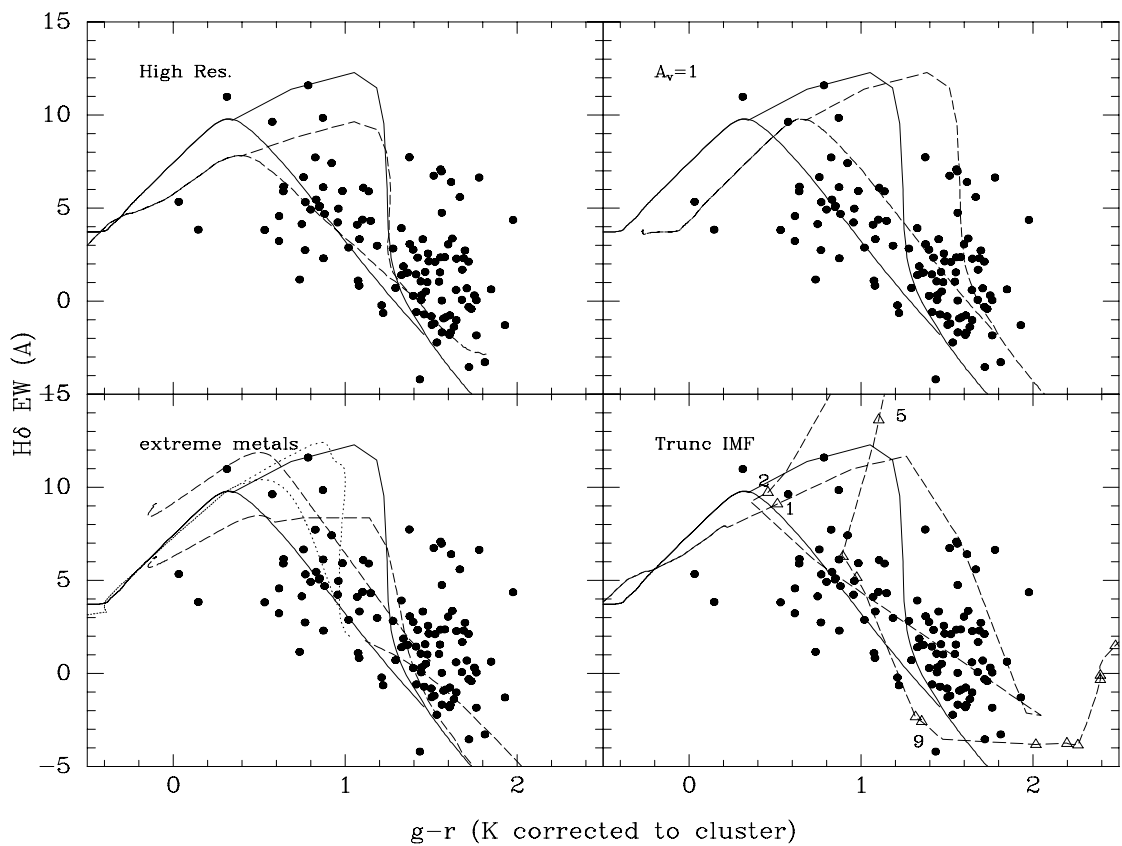}{4in}{90degrees}{75}{75}{288pt}{-36pt} 
\caption[fig18]{As Figure~\ref{fig17} with the solar metallicity
standard models plotted as solid lines, but also with model lines
plotted for: (top-left) a Bruzual and Charlot solar metallicity model
from 1995 with much higher spectral resolution, (top-right) the
standard model shifted for a reddening A$_{\rm v}$=1 (E(g-r)=0.3312),
(bottom-left) 5 times solar metallicity (dashed line), and 1/50 times
solar metallicity (dotted line), and (bottom-right) Bruzual and Charlot
solar metallicity model from 1995 with an IMF with no stars of mass
less than 2.5 times solar. A few times (in Gyr) have been marked on the
model line. For clarity, the error bars have been removed from the data
points. The error bars can be seen in Figure~\ref{fig12}.  
\label{fig18}}
\end{figure}

\clearpage

\begin{figure} 
\plotfiddle{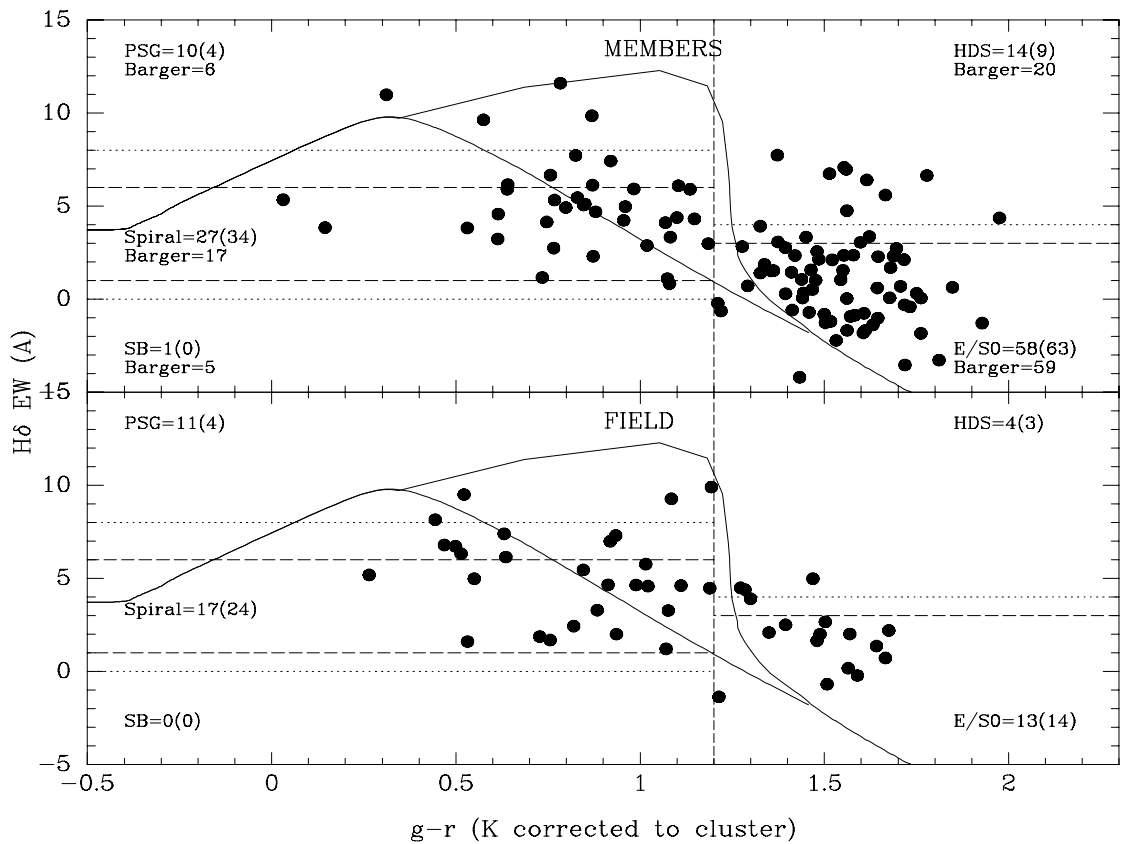}{4in}{90degrees}{75}{75}{288pt}{-36pt} 
\caption[fig19]{As Figure~\ref{fig17} with the solar metallicity
standard models plotted as solid lines, but also with the dividing
lines proposed by \cite{bar96} shown as dashed lines, and some
alternative dividing lines proposed in this paper shown as dotted
lines. Top panel shows cluster members, bottom panel shows the field
sample for the same redshift. See the text for explanation of the
numbers listed. For clarity, the error bars have been removed from the
data points. The error bars can be seen in Figure~\ref{fig12} and
\ref{fig13}.
\label{fig19}}
\end{figure}

\clearpage

\begin{figure}
\plotfiddle{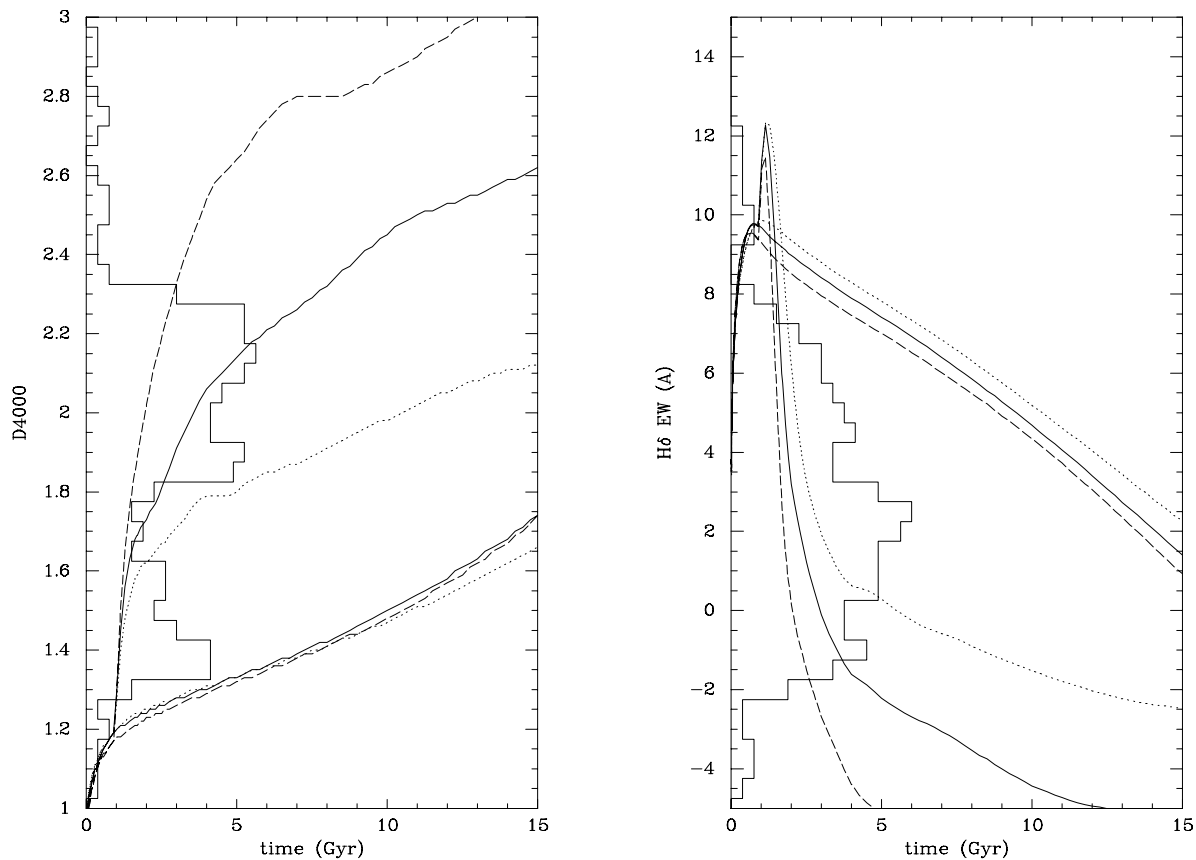}{4 in}{90degrees}{75}{75}{288pt}{-36pt}
\caption[fig20.eps]{Age/Metallicity measures of the oldest galaxies.
Left panel shows model curves of D4000 versus age from solar (solid),
2.5 times solar (dashed) and 0.4 times solar (dotted) metallicity
models. The upper curves are for a 1 Gyr burst of star formation
(ellipticals), while the lower curves are for exponentially falling
star formation with a time constant of 4 Gyrs (spirals). Also plotted
is the histogram of observed D4000 for cluster members. The Right panel
shows the same curves and data but for H$\delta$ rather than D4000. The
`spiral' curves in this case lie above the `elliptical' curves for
times later than 3 Gyrs.
\label{fig20}}
\end{figure}

\end{document}